\documentclass[preprint2]{aastex}

\usepackage{graphicx}
\usepackage{amssymb}
\usepackage{epstopdf}
\begin{document}

\title{Spectropolarimetric observations of Herbig Ae/Be Stars I: HiVIS spectropolarimetric calibration and reduction techniques.}
\author{D. M. Harrington \& J.R. Kuhn}
\affil{Institute for Astronomy, University of Hawaii 2680 Woodlawn Drive, Honolulu, HI, 96822 }
\email{dmh@ifa.hawaii.edu}
\email{kuhn@ifa.hawaii.edu}

\begin{abstract}

	Using the HiVIS spectropolarimeter built for the Haleakala 3.7m AEOS telescope in Hawaii, we are collecting a large number of high precision spectropolarimetrc observations of stars.  In order to precisely measure very small polarization changes, we have performed a number of polarization calibration techniques on the AEOS telescope and HiVIS spectrograph.  We have extended our dedicated IDL reduction package and have performed some hardware upgrades to the instrument.  We have also used the ESPaDOnS spectropolarimeter on CFHT to verify the HiVIS results with back-to-back observations of MWC 361 and HD163296.  Comparision of this and other HiVIS data with stellar observations from the ISIS and WW spectropolarimeters in the literature further shows the usefulness of this instrument.

\end{abstract}


\section{Introduction}

	High-resolution linear spectropolarimetry measures the change in linear polarization across a spectral line.  Many models show this measurement is a useful probe of circumstellar environments at small spatial scales.  Circumstellar disks, rotationally distorted winds, magnetic fields, asymmetric radiation fields (optical pumping), and in general, any scattering asymmetry can produce a change in linear polarization across a spectral line such as $H\alpha$ (cf. McLean 1979, Wood Brown \& Fox 1993, Harries 2000, Ignace et al. 2004, Vink et al. 2005a, Kuhn et al. 2007).  These signatures can directly constrain the density and geometry of the circumstellar material and probe the near-star environment.  Typical spectropolarimetric signals are small, often a few tenths of a percent change in polarization across a spectral line.  Measuring these signals requires very high signal to noise observations and careful control of systematics to measure signals at the 0.1\% level.  		

	This technique probes small spatial scales, typically being sensitive to the region within 10 stellar radii.  Even for the closest young stars (150pc), these spatial scales are smaller than 0.1 milliarcseconds across and will not be imaged directly, even by 100m telescopes.  Since the circumstellar material is involved in accretion, outflows, winds and disks, with many of these phenomena happening simultaneously, spectropolarimetry can put unique constraints on the types of densities and geometries of the material involved in these processes.  

	In order to address these issues, we built a spectropolarimeter for the HiVIS spectrograph on the 3.7m AEOS telescope and performed a preliminary telescope polarization calibration (Thornton 2002, Thornton et al. 2003, Harrington et al. 2006).  Since then, we have made many improvements.  We have upgraded much of the hardware, calibrations, and reduction package.  A large number of stellar observations have been partially reported in Harrington \& Kuhn 2007 to illustrate the capability of the spectropolarimeter.  We present the reduction package, polarization calibrations, and implications of these calibrations for stellar observations with this instrument.

\section{Improvements and A Dedicated Reduction Package}

	We have done several upgrades to the instrument recently.  The polarimeter has been remounted with better alignment and less vignetting.  A new tip-tilt guiding corrector has been installed and calibrated.  A hard-mounted, adjustable dekkar and polarization calibration mount have been installed and allow for quick, easy, and repeatable calibration and adjustment.  The failed CCD-ID20's were replaced with a similar copy, but an unresolved hardware/RFI issue caused this repair to fail.  The original science camera failed and has been replaced by a new commercial 1k$^2$ PIXIS camera and is working wonderfully.  The calibration stage had lenses mounted that increased the flat-field flux by a factor of 6 and the ThAr lamp flux by a factor of 150.  These improvements greatly increase the ease and accuracy of spectral calibration.  

  The spectropolarimetric reduction pipeline can be broken into a few basic parts.  1) Combining flats, darks, biases. 2) setting up the coordinate grids as x-pixel, y-pixel, slit-tilt and wavelength calibration.  3) Calculating a first-pass spectrum by simply averaging all pixels at a given wavelength.  4) Calculating more spectra based on optimal-extraction routines, with cosmic ray rejection and flux-weighting.  5) Doing the polarization calculations and 6) Simulating and quantifying the systematic errors.

	\subsection{Combining flats and setting up the coordinate grid}

  The first step is trivial - a clipped mean is performed on a number of flats to produce a flat field.  Since we are doing high S/N measurements, we found that we needed a minimum of 20 flats at 10000 ADU to get good S/N.  We decided to do the dark and bias calibration in a less-than-traditional way because of a slowly-fluctuating, non-repeatable background level, which is quite common with most detectors.  We had an overscan region in the original science detector with the CCD-ID20's which allowed us to track the changing dark levels throughout the night.  The replacement Pixis 1024BR array was smaller than the focal plane and is always exposed to light.  Both the original Lincoln Labs CCD-ID20's and the Pixis detector showed a significant ($>$few\%) drift in the bias levels throughout a night.  The Pixis camera had no overscan region, but the median values of all the darks taken over many months did show variation in the overall background levels.  These variations with this instrumental setup and background levels required a background subtraction from each frame using the unexposed pixels outside the echelle orders as this directly tracked the background levels of the chip and reduced the systematic uncertainties.  For our short exposures, the dark current was small, and is automatically corrected with the frame-by-frame background subtraction.
  
\begin{figure}[!h!t!b!p]
\includegraphics[width=0.7\linewidth, height=1.0\linewidth, angle=90]{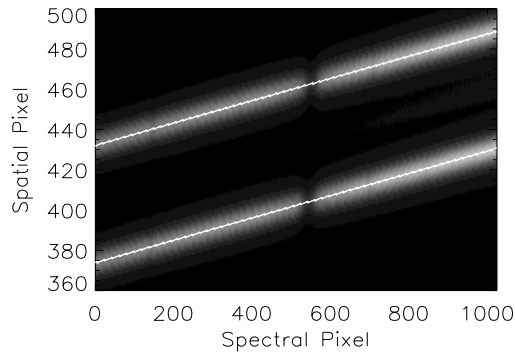}
\figcaption[coord]{\label{fig:coord} The user-defined points serve as a starting point for the peak-finding routine that fits a 3$^{rd}$ order polynomial to the shape of the order, as specified by a bright stellar template.  The gaussian-fit centers are plotted as the jagged line.  The polynomial fits to the peak intensity follow the stellar brightness peak quite nicely.}
\end{figure}

  Since the instrumental setup and the order locations on the CCD's has changed much over the last few years, and changed quite often during the instrument work done in the fall of 2006, setting up the coordinate grid requires more user-input than most dedicated reduction scripts.  The user must supply a guess at the polarized order locations, typically 5 points to fit a 3rd-order polynomial, and a bright stellar image taken that night to serve as a template for the curvature of the orders.  Figure \ref{fig:coord} shows a bright standard with the narrow H$\alpha$ line near the center of the 1k$^2$ array.   
  
 %%%%%%%%% Figure 1 here

  The code sets up an initial polynomial fit to these user-input points, and then takes spatial (vertical) cuts at each spectral (horizontal) pixel. Each of these spatial cuts are then fit by a gaussian to determine a robust location of the peak intensity in the spatial direction across the CCD, reducing the influence of cosmic rays or chip defects.  These peak intensity positions allow for a robust fit to the order shape on the CCD with a slope of around 5\% (crossing a spatial pixel every 18 or 20 spectral pixels).  The orders are fit independently and the shapes are the same to the 0.1 pixel level ($\sim$0.2\% of an order width).  The order separation, caused by the Savart plate, is then calculated as the median difference between the two polynomial fits.  For the Pixis camera, this was 58-59 pixels.  The residuals between the calculated peak intensity and the polynomial fits are typically under half a pixel and show the expected oscillation with wavelength as spatial pixels are crossed.  
  
\begin{figure}[!h!t!b!p]
\includegraphics[width=0.7\linewidth, height=1.0\linewidth, angle=90]{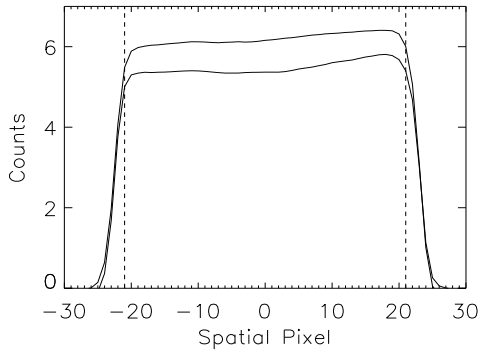}
\figcaption[flatslice]{\label{fig:flatslice} The average spatial profile of the two polarized flat-field orders across the H$\alpha$ region.  The width of the orders is set by by the dekkar width.  The vertical lines mark the edge of the illuminated field we selected for use.  The small tilt of the spatial cuts traces the inhomogenity of the illumination field at the slit.  The difference in height between the two cuts corresponds to the intrinsic continuum polarization of the instrument downstream of the calibration stage.}
\end{figure}

     The star observed for the template is never completely centered in the slit from small guiding errors.  Once we have these order shapes from the stellar observations, we then must adjust these polynomials to reflect the true slit center.  This is done with the flat-field.  First we create an average spatial profile (slit illumination pattern) for the flat field, centered on the stellar template by simply averaging the flat field along the order template in the spectral direction.  These spatial profiles, shown in figure \ref{fig:flatslice} clearly show the slit boundaries, where the flat field intensity falls off sharply.  The flat-field spatial cuts are cross-correlated to find the offsets from the stellar order-shape polynomial fits to the center of the flat-field spatial cuts.     

 %%%%%%%%% Figure 2 here

  Once the offset from the stellar template to the flat-field region is calculated, the full coordinate grid can be computed with the order centers, edges, and background locations specified.  Once the coordinate grid is set, we can then make a calibrated flat field, adjusting for the blaze function, and wavelength dependence of the flat-field lamp (blackbody). 

   A flat-field is supposed to correct for pixel-to-pixel gain and field-dependent efficiencies only.  The flat-field values should not introduce any blaze, lamp or  spatial illumination dependencies.  This means that the flat-field must be corrected for it's inherent spatial and spectral structure.  To correct the flat field spectral dependence (horizontal), a simple flat-field spectrum is made for each polarized order by averaging all pixels spatially at a given spectral position.  This spectrum has a bit of intrinsic noise in it so a low-order polynomial fit is made.  The flat-field image then has this spectral dependence removed.  

 %%%%%%%%% Figure 3 here

\begin{figure}[!h!t!b!p]
\includegraphics[width=0.7\linewidth, height=1.0\linewidth, angle=90]{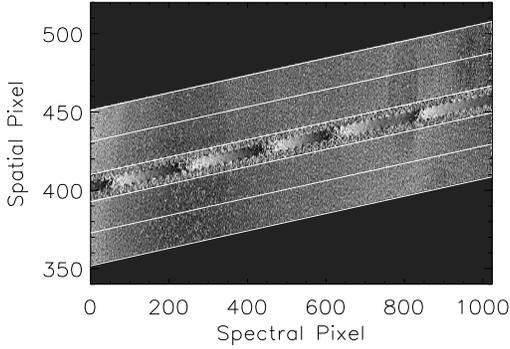}
\figcaption[calibratedflat]{\label{fig:calibratedflat} The calibrated flat field showing the center of each order and the width of each order. The region used for the background subtraction begins 10 pixels above and below the order.  The grey-scale runs from 0.95 to 1.05 showing the pixel-to-pixel structure.  The region in the middle is excluded from the reduction.}
\end{figure}

  The spatial structure is corrected by dividing each spatial cut along the flat-field spectrum by the average spatial profile, removing the spatial dependence of the flat (inhomogeneous slit illumination) at each wavelength.  This makes our flat-field much closer to a true pixel-to-pixel gain measurement, shown in figure \ref{fig:calibratedflat}.

	\subsection{Wavelength Calibration}

  With the recent calibration-stage improvements we have implemented, it became possible to do very accurate wavelength calibration using low order polynomial fits to at least six clear, unblended lines, just over the H$\alpha$ spectral region.  With a $>$600x increase in ThAr flux (15x without the diffuser-bypass mirror), we can use the smallest slit (0.35") at highest resolution (R=49000) to clearly identify lines and compare the lines with standard line-lists obtained from the NOAO spectral atlas.  The new HiVIS detector has a wavelength scale of 0.037$\AA$ per pixel (1.7 km/sec per pixel at H$\alpha$).  Thus the rotation of the Earth produces a $\frac{1}{3}$ pixel shift and seasonal variations can approach 40 pixels.  
  
 %%%%%%%%% Figure 4 here

\begin{figure}[!h!t!b!p]
\includegraphics[width=0.8\linewidth, height=1.0\linewidth, angle=90]{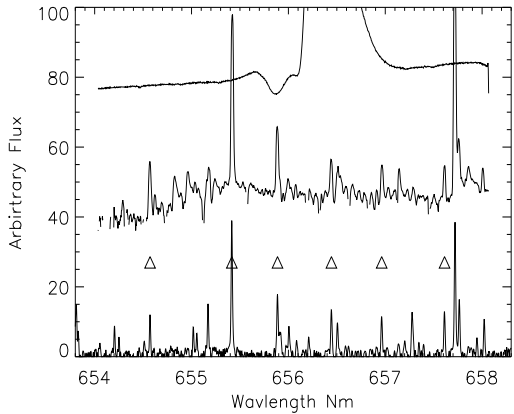}
\figcaption[thar-fit]{\label{fig:thar-fit} The NOAO ThAr spectra is on the bottom.  The calculated HiVIS 0.35" ThAr spectrum is in the middle.  The triangle symbols mark the lines used in the polynomial fit.  A sample unprocessed AB Auriga spectra is shown as the top line to illustrate the wavelength coverage of a single spectral order.}
\end{figure}

  The wavelength calibration is done in a two-step process.  A Th-Ar spectrum taken with the 0.35" slit is used to identify the lines.  The Th-Ar spectra are calculated with the simple average over all illuminated pixels.  The unblended lines are identified and the peaks are fit by a gaussian to determine the line centers.  A 3$^{rd}$ order polyomial fit to the line centers is used to create a wavelength array.  The NOAO atlas spectrum, HiVIS spectrum, and a comparison AB-Auriga spectrum are shown in figure \ref{fig:thar-fit}.  The triangles mark the 6 lines used in the polynomial fits.  Since all stellar observations were done with the 1.5" slit, a 4-orientation sequence of ThAr spectra are taken with this slit and then reduced with the simple-average method.  The resulting lower-resolution spectrum is cross-correlated with the higher-resolution spectra produced by the 0.35" slit to determine any shifts between slits and waveplate orientations, the beam wander or wobble, which is a small fraction of a pixel.  

 %%%%%%%%% Figure 5 here

\begin{figure}[!h!t!b!p]
\includegraphics[width=0.7\linewidth,height=1.0\linewidth, angle=90]{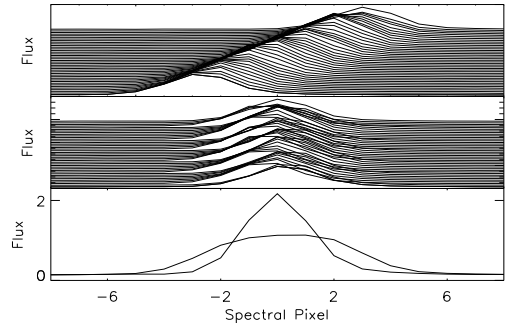}
\figcaption[thar-align]{\label{fig:thar-align} The spectral shift of the Th-Ar lines along the slit is calculated and used to set up a tilted coordinate grid.  Main result is a 70\% increase in resolution between reductions done with tilted and untilted coordinate grids (R=30000 to R=49000 for the 0.35" slit).  }
\end{figure}

    The average slit tilt is then calculated to modify the coordinate-grid for the reduction program.  The tilt is a spectral-direction shift of $\sim$7 pixels between the bottom and top of each order (across the 50 pixel slit width).  This shift is calculated as a cross-correlation of a spectral cut beween each spatial position along the order.  This tilt is then implemented as a integer-pixel shift of the coordinate grid at each spatial point.  There is a very significant increase in resolution when the tilted-coordinates are used.  Figure \ref{fig:thar-align} shows the spectral cuts before and after alignment, as well as the intensity of the line calculated with both tilted and un-tilted coordinate grids.  The spectral resolution derived from the Th-Ar line fits goes from R=30000 to R=49000 when using tilted coordinates. 

	\subsection{Calculating Raw Spectra}

  Once the tilted coordinate grid, gain-table, and wavelength calibration is done, the data can be processed.  Each exposure is divided by the gain-table (calibrated flat). A background spectrum is made by averaging the unexposed background region outside the polarized orders.  This background is noisy and so a linear fit to this background is subtracted from each order to remove the dark current and bias of the detector.  
  
  To create a simple, first-pass at the spectrum, every pixel at each wavelength was simply averaged to compute the intensity.  At the same time, an average spatial profile (PSF) is constructed by averaging all these spatial cuts spectrally.  While this method does not correct for cosmic rays, and does not weight the pixels by their fractional-stellar-flux or PSF, this simple average spectrum provides a starting point for more sophisticated reductions.
   
  	\subsection{Shift-n-Scale}
  
       In many parts of the reduction package, we want to shift and scale one curve to another.  We have developed a least-squares routine that allows us to calculate the shift and scaling coefficient between two input curves.  The routine assumes that the scale and shift from one curve to another can be written in terms of the input curve and it's derivative.  To illustrate the routine we will show a shift-n-scale between the two polarized spectra, $I_1$ shifted and scaled to $I_2$, as below:

	\begin{equation}
	I_2 = aI_1(x-b) 
	\end{equation}

	I is the intensity as a function of CCD pixel x, a is the relative gain between the two orthoganally polarized spectra, usually $\sim$1, and b is the spectral shift in pixels, usually $\leq$0.8 pixels.  The spectrum $I_1(x-b)$ can be expanded to first order as below:

	\begin{equation}
	 I_2=aI_1- ab(dI_1/dx)  
	\end{equation}

	Rewriting the constants $a=K_1$ and $-ab=K_2$ and bringing all terms to one side gives the difference between the spectra $I_2$ and $I_1$ scaled and shifted by a and b.  The least squares routine minimizes the function below for all pixels (x):

	\begin{equation}
	 \Sigma (I_2-K_1I_1-K_2dI_1/dx)^2.  
	\end{equation}

 %%%%%%%%% Figure 6 here

\begin{figure}[!h!t!b!p]
\includegraphics[width=0.7\linewidth, height=1.0\linewidth, angle=90]{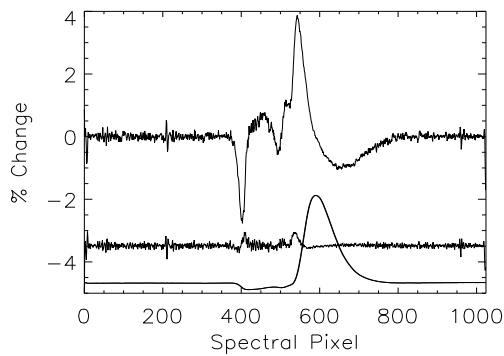}
\figcaption[shfscl]{\label{fig:shfscl} An example of the shift-n-scale algorithm correcting for optical misalignments.  A single spectrum, the bottom curve, was replicated and shifted spectrally by 1 pixel and then fed to the shift-n-scale routine.  The top curve is the resulting large-amplitude (5\% $\frac{a-b}{a+b}$) fractional differences.  This curve follows the derivative and must be corrected.  The shift-n-scale routine gave a scale of 0.999846 and a shift of 1.00742.  The shift-n-scale interpolation causes some residual noise, since the input curves were identical and should have subtracted completely.  This arises because of the smoothing done to calculate the derivative.  The middle, mostly flat curve shows this residual noise doubled for clarity.  The noise is at the level of 0.2\% with some residual systematic error.}
\end{figure}  

	The coefficients $K_1$ and $K_2$ are then used to calculate the scaled and shifted functions.  An example of this routine applied to a spectrum of AB Aurigae is shown in figure \ref{fig:shfscl}.  A single spectrum was replicated and shifted by one pixel and corrected by the shift-n-scale routine.  The routine uses a smoothed profile to calculate the derivative.  This illustrates how a misalignment of the savart plate, simulated as a spectral shift of 1 pixel between the orthogonally polarized spectra, produces a fractional difference of order 5\% from this misalignment.  This signature is a systematic error that follows the derivative of the spectral line.  The scale and shift operation minimizes this derivative signal as shown (amplified) in the bottom line in the plot.  The residual noise is at the level of 0.2\% and has some residual systematic error in it from the smoothing used to calculate the derivative, but the derivative-based error has been attenuated by a factor of about 20.

	The shift-n-scale routine also has some nonlinearity to it, and is best used for pixel shifts $<$0.8 pixels.  When shift-n-scaling the original curve to the simulated one, the routine reproduced the shifts to better than 0.05\% for spectral shifts of 5 pixels, far greater than the range needed.  However, when shift-n-scaling the simulated curves to the original curve, the calculated shifts had an error of $>$0.1 pixels for spectral shifts $>$0.8 pixels.  This sets the useful range of this routine.

 %%%%%%%%% Figure 7 here
  
\begin{figure}[!h!t!b!p]
\includegraphics[width=0.8\linewidth, height=1.0\linewidth, angle=90]{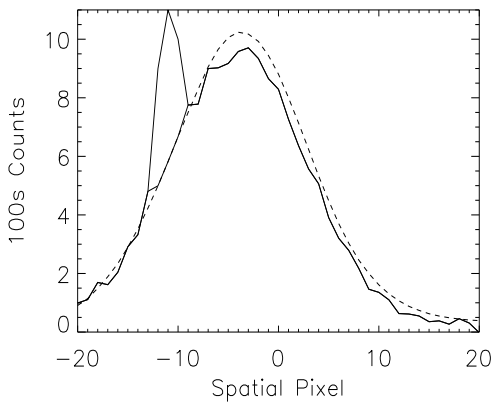}
\figcaption[cosray]{\label{fig:cosray} This shows a cosmic-ray removal from a spatial cut.  The cosmic-ray removal routine computes the deviation of the spatial cut from the average line profile and identifies 4-$\sigma$ deviations in each spatial cut from the average spatial-profile (PSF) and replaces  the cosmic ray hits with the value for the least-squares scaled average spatial profile.  The smooth dashed curve is the computed least-squares scaled average spatial profile and the jagged curve is an individual spatial cut.  Three pixels were found to be $>4\sigma$ away from the average profile and were corrected.}
\end{figure}

	\subsection{Optimal Spectral Extraction}

  This first-pass spectrum and average spatial profile are projected back on the original spatial cuts with the shift-n-scale routine.  The expected noise at each spatial pixel is found (using estimates of the gain and read-noise).  The standard deviation of each spatial pixel at each wavelength is then calculated and used to find any cosmic-ray hits as cosmic-rays (and chip defects) which will be strong devations ($>4\sigma$) from the spatial profile.  An example of this cosmic ray removal routine is shown in figure \ref{fig:cosray}.  The average spatial profile is the dashed line.  The spatial cut at this wavelength shows a few pixels deviating strongly from the average spatial profile from a cosmic ray hit.  These pixels are identified and replaced with the shifted-and-scaled spatial profile, greatly reducing the error in the spectrum.  Once this cosmic ray removal routine is done, a spectrum is calculated as the simple average of all the pixels.  The corrected spatial cuts are passed on to more optimal spectral extraction routines.
  
    The first of the optimal extraction routines uses the shift-and-scale routine to fit the average spatial profile (PSF) to each of the individual spatial cuts at each wavelength.  This reduction process assumes that the spatial profile does not vary across the order, an assumption that all optimal reduction packages adopt (Horne 1986, Marsh 1989, Donati et al. 1997, Cushing et al. 2004).  The calculated error in the shifts between the optional fit and the calculated order location is dependent on the flux, but is typically less than 0.1 pixels.
        
    Another set of spectra are calculated as simple gaussian fits to the spatial cuts.  While the PSF (spatial cut) is not entirely gaussian, and certain seeing conditions or tracking-errors lead to not-quite-gaussian spatial cuts (PSF's), the fits are usually robust, and match the shift-n-scale and simple-average specta well.  The standard gaussfit routine is applied to each spatial cut, with the initial estimates for the gaussian as:  peak intensity is twice the computed average-spectra, centered on the order, with a fwhm of one-quarter the order width and zero background.  The routine works well when the spatial cuts (PSF's) are roughly gaussian.
    
 %%%%%%%%% Figure 8 here
	
\begin{figure}[!h!t!b]
\hspace{-5mm}
\includegraphics[width=0.9\linewidth, height=1.2\linewidth, angle=90]{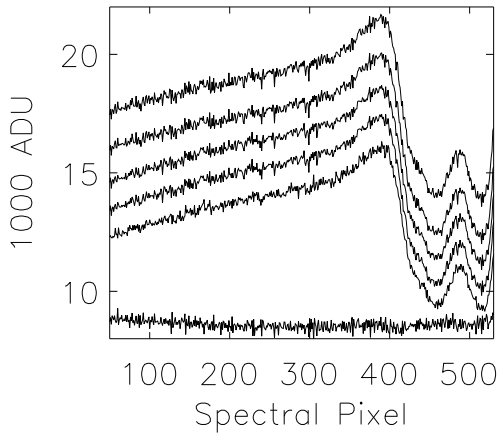}
\figcaption[difred]{\label{fig:difred} This shows a spectrum of AB Aurigae reduced with varying parameters to show the robustness of the reduction program.  Each curve represents a different reduction using order widths from 43 to 49, with or without flat fielding, and with or without tilted-coordinates.  From top to bottom 1) tilted, 49 width, no flat 2) tilted, 43 width, no flat 3) tilted, 43 width flattened 4) tilted 45 width flattened  5) untilted 49 width no flat.  The bottom curve shows the residuals from the unflattened untilted 49-width curve and the flattened tilted 43-width curve.}
\end{figure}
	
        Thus, we have three spectra with different systematic effects (simple-average, shift-n-scale, gaussian-fit) -  which are all included in the polarization analysis to probe our polarimetric reduction codes and systematic errors inherent in these reductions.  In practice, the exact choice of reduction parameters is unimportant.  Figure \ref{fig:difred} shows the simple-average method with and without a flat field or tilted coordinates, and varying order widths from 43 to 49 pixels (covering the edges of the illuminated region, see figure \ref{fig:flatslice}).  The tilted coordinates do add a significant increase in resolution, but for the high signal-to-noise spectropolarimetric observations the spectra are rebinned to lower resolution anyways.  The flat field does increase the signal-to-noise of the intensity calculation, but as we show below, the basic polarization calculations are independent of the flat field.  
        
        The AEOS telescope is designed for military applications and does not have a standard siderial guiding telescope.  There are significant tracking errors that can accumulate and there are offsets induced by manual telescope-operator corrections.  The operation requirements set out by the AFRL require that we have no control over the guiding, and that their operators control the pointing of the telescope.  There are a number of data set's where an operator's manual steps caused a non-gaussian spatial profile.  Thus we calculate the spectra using all three methods, but the averaging and shift-n-scaling are most robust. 

  	These three different reductions vary slightly between each other.  The gaussian-fit and simple-average methods vary by only a few parts-per-million when the spatial cuts (psf's) are gaussian.  However, the shift-n-scale method and the simple-average methods vary slightly across the line profile, giving an overall uncertainty in the shape of the intensity profile.  However, the spectropolarimetry is calculated as the difference between line profiles on the same image.  Since the same reduction method is applied to each polarized order in the same way for each polarization measurement, it is only any differential error that would cause systematic errors in the spectropolarimetry.  The reduction package uses all three calculation methods to assess the systematic errors that might be present in each method, or in the subsequent processing of the spectra.
   
	\subsection{Continuum Subtraction}
  
  Our focus is on spectral and spectropolarimetric variability.  With the many instrumental polarization effects, we must do continuum-polarization subtraction to isolate the change across the line profile.  Since spectral variability is always referenced to continuum, no spectrophotometric or flux calibration is done, though standards are observed for future work.  
  
  The continuum is simply subtracted as a linear fit to the intensity on either side of the spectral line.  This continuum subtraction normalizes the slope of the black-body of the star, removes the instrumental continuum polarization signals, and removes the instrumental blaze efficiency.  Continuum subtraction is done by averaging 20 pixels sufficiently far enough away from chip edges.  We found that there is a variable dark signal that curves the echelle orders near the edge of the chip, 10-20 pixels wide, in a non-repeatable way, so this region is avoided.  Cirrus, tracking errors, and wind-bounce all influence the throughput of the spectrograph and thus, even on a spectrophotometric night, a significant variation in the continuum flux is sometimes seen.
  
    The data taken before 10-27-2006 was imaged using an echelle order which placed the H$\alpha$ line near the edge of the chip.  This placement was done before we did some instrumental corrections, and before we remounted the Pixis camera to a different region of the focal plane.  In the more recent data, H$\alpha$ is closer to the blaze-efficiency peak and is much closer to the center of the focal plane.  The continuum fits to the CCD-ID20 images were not as accurate because the red wings of the H$\alpha$ line were near the order edge, giving rise to a small ($\sim$1\%) error in slope estimate.  However, this uncertainty is negligable for data taken on each night because the continuum is calculated in the same way for each instrumental setup on each night.  This only produces an uncertainty in the night-to-night variability of the line profiles, and in the overall slope of the calculated continuum and continuum polarization.

\subsection{Possibility of Spectro-astrometry}

 	At this point it is useful to demonstrate the coordinate grid and basic properties of the reduction package by demonstrating the wavelength dependence of the point-spread function properties.  Several authors have argued that the change in centroid location (brightness peak) or full-width at half-max (fwhm) can be used as evidence for binarity in a set of spectroastrometric calculations (Beckers 1982, Bailey 1998, Baines et al. 2006).  Spectroastrometry in principle is a technique that can detect unresolved binaries as a change in spatial profile width (psf) and location (centroid) across an emission or absorption line where the relative brightness of the two binary components significantly varies across the line.  For example, if one star has an emission line and the other an absorption line, the centroid will shift towards the emitter in the core of the emission line.  If the binary is significantly wide and bright, the fwhm would also decrease across the line as the relative contribution to the psf from the wide companion is reduced.  This method makes use of the assumption that the psf across a line or spectral order does not change much, and that any significant change can be attributed to the source.  HiVIS has been used to observe a number of young or emission-line stars as part of a large survey (Harrington \& Kuhn 2007) all of which have strong H$\alpha$ lines.  Given the reported detections, we will present the feasibility and capability of HiVIS to do similar work. 

 %%%%%%%%% Figure 9 here

\begin{figure}[!h!t!b!p]
\hspace{4mm}
\includegraphics[width=0.75\linewidth, height=0.95\linewidth, angle=90]{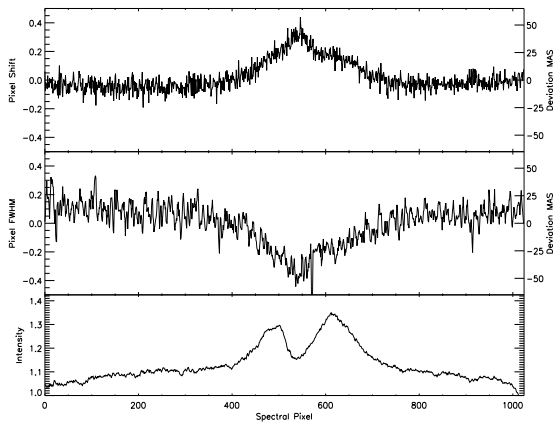}
\figcaption[specast-mwc166]{\label{fig:specast-mwc166} The spectroastrometry of MWC 166.  The top box shows the center of a Gaussian fit to the PSF at every spectral pixel.  The middle box shows the calculated FWHM for each spectral pixel.  The bottom box shows the resulting normalized intensity.  A systematic effect is seen across the line. }
\end{figure}  
	
	Binaries are common among pre-main-sequence stars and this technique can easily be applied to every data set taken with this instrument.  It also provides a good check on any systematic errors such as order mis-fitting or psf mis-fitting since any systematic error will show up with this measurement technique.  Since the centroid of a spatial profile on most spectrographs can be calculated to a fraction of a pixel accurately in high signal-to-noise data, this technique has the capability to detect binarity to less than 30mas separations or a third of a pixel when our pixels are 130mas (cf. Bailey 1998, Baines et al. 2006).  Normally, with a long-slit spectrograph, the slit of the instrument is rotated to a north-south line and an east-west line to fully sample the spatial changes.  This allows one to establish the position-angle of the binary system.  For polarimetric reasons expanded below, we do not use the image rotator and keep this component fixed vertically during our observations.  This means that the position-angle rotates from exposure to exposure, but presenting the calculations is still useful to demonstrate the capability of the analysis software. 

 %%%%%%%%% Figure 10 here

\begin{figure}[!h!t!b!p]
\hspace{3mm}
\includegraphics[width=0.75\linewidth, height=0.95\linewidth, angle=90]{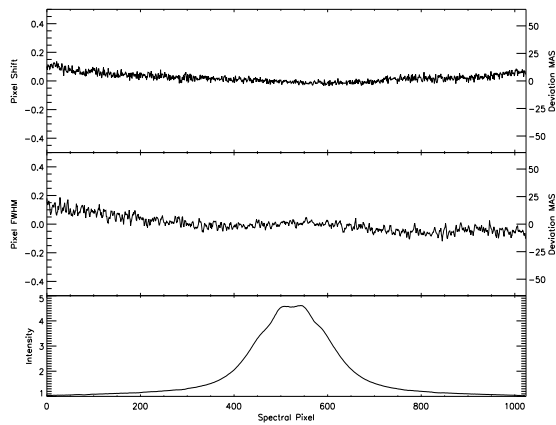}
\figcaption[specast-gmcas]{\label{fig:specast-gmcas} The spectroastrometry of $\gamma$ Cassiopeia.  The top box shows the center of a Gaussian fit to the PSF at every spectral pixel.  The middle box shows the calculated FWHM for each spectral pixel.  The bottom box shows the resulting normalized intensity.  No systematic effect is seen across the line. }
\end{figure}  

 %%%%%%%%% Figure 11 here	

\begin{figure}[!h!t!b!p]
\vspace{-2mm}
\hspace{2mm}
\includegraphics[width=0.75\linewidth, height=0.95\linewidth, angle=90]{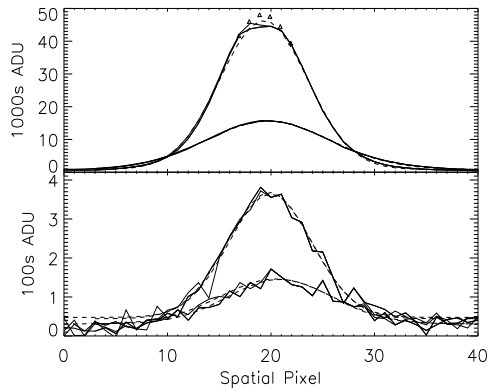}
\vspace{-5mm}
\figcaption[specast-pcyg-psf]{\label{fig:specast-pcyg-psf} The spatial profile or psf of P-Cygni on two separate exposures and three different reductions.  One exposure was clipped (nearly saturated) to illustrate a systematic error in wavelength dependence of the psf.  The top panel shows the psf at the emission peak.  The triangles mark a gaussian fit to the non-clipped part of the psf.  The bottom panel is the psf in the absorptive trough.  The dashed lines in both panels show the Gaussian fits.  Notice how the wings of the Gaussian are significantly above the noise in the low-flux absorptive trough.}
\end{figure}  
	
	For each spectrum, the optimal spectral extraction algorithms fit Gaussians and average spatial profiles (psf's) to the data at each wavelength, as was shown in figure \ref{fig:cosray}.  These spatial profiles can easily be compared with the nominal values for the average spatial profile's FWHM and centroid for each order.  At each wavelength, a deviation for the FWHM and the centroid can be calculated.  Because of small guiding errors, the centroid of each stellar observation does not coincide with the exact order center.  Because of the tilt of the orders on the detector, this will cause the centroid of the spectrum to cross spatial pixels at different spectral pixel locations.  In order to correct for this, our analysis package does a cross-correlation between the nominal order center location, as illustrated by the jig-saw shaped fit in figure \ref{fig:coord}, and the centroid location of the actual observed spectrum.  Once this small offset has been corrected, the centroid location and the FWHM across the order is calculated.  We will discuss some sample calculations from data taken the 20$^{th}$ of September, 2007 below. 
 
	In Baines et al. 2006, the known binary MWC 166 was reported to have a centroid shift of roughly 40mas EW and 15mas NW with corresponding FWHM changes of 40mas and 5mas respectively.  This is tabulated in their paper as a total shift of 50mas at a PA of 287$^\circ$ where the known separation of the binary is 0.65" at a PA of 298$^\circ$.  Figure \ref{fig:specast-mwc166} shows a HiVIS observation of this star.  The observations had a peak of 1500ADU and a continuum of 800-1300ADU across the order and a fwhm of 11.7 pixels (1.5").  The deviation we detect in FWHM and centroid offset are both around 40mas, both the correct magnitude for this system.  
	
	We also present observations of $\gamma$ Cassiopeia in figure \ref{fig:specast-gmcas} at the same scale as MWC 166 to show a non-detection across a strong line.  The peak intensity was near 26000ADU with continuum from 4000-6000ADU and a fwhm of 10.1 pixels (1.3").  This non-detection shows that spurious signatures caused from the 1\% Savart plate leak, non-linearity in the chip, spatial dependence of the focus or any other effects are not present across an emission line. 

 %%%%%%%%% Figure 12 here

\begin{figure}[!h!t!b!p]
\hspace{4mm}
\includegraphics[width=0.75\linewidth, height=0.9\linewidth, angle=90]{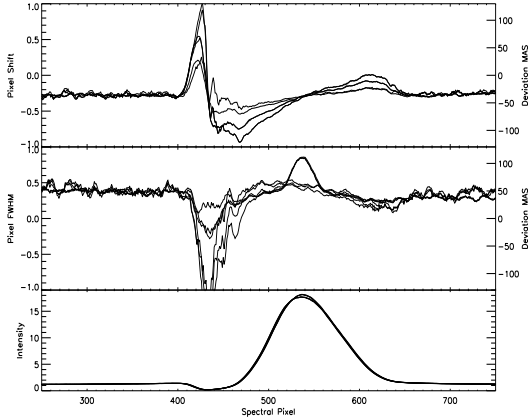}
\figcaption[specast-pcyg]{\label{fig:specast-pcyg} The wavelength dependent psf properties of P-Cygni for two separate exposures and three separate reductions.  The three different reductions are tilted and untilted coordinates with width of 47 pixels, and untilted coordinates with a width of 41 pixels.  The top box shows the center of a Gaussian fit to the PSF at every spectral pixel.  The saturated exposure shows a smaller deviation than the unsaturated exposure for every type of reduction meaning that more flux gives better fits.  The tilted coordinates do not change the result.  The middle box shows the calculated FWHM for each spectral pixel.  The saturated exposure shows a 50mas increase in fwhm in the clipped part of the line.  The unsaturated exposure shows a much greater 200mas decrease in fwhm in the absorptive component of the line.  The tilted coordinates do not change the result, but using a smaller order width decreases the deviation, again suggesting that greater flux influences the fits.  The bottom box shows the resulting normalized intensity.  A systematic saturation effect is seen for all reductions across the emission peak of line where the saturated exposure is underestimated. }
\end{figure}  

	The star P-Cygni has a very strong H$\alpha$ line.  We observed this star with two different exposure times separated by 90 minutes.  The reduction was performed with three separate settings: a width of 47 for tilted and untilted coordinates, as well as a width of 41 un-tilted.  This will be used to highlight changes in derived psf properties when including tilts and neglecting low-flux psf wings.  The altitude changed from 50$^\circ$ to 34$^\circ$ and the frame rotated by 28$^\circ$ between the observations.  The line has a normalized intensity of 0.194 in the absorption trough and 18.180 in the emission peak, or a flux ratio of 94.  This huge range in flux and the difference in seeing serves to highlight systematics.  One exposure is deliberately clipped (near saturated) to illustrate systematic errors that can occur with this technique.  The spatial profiles are shown in figure \ref{fig:specast-pcyg-psf}.  The peak intensity of the saturated exposure was 44000ADU with continuum of 1800-3800ADU and only 400ADU in the absorptive component of the line. The calculated fwhm was 9.6 pixels (1.25").  However, a Gaussian fit to the unsaturated region shows that the true peak should have been closer to 48000ADU, or a roughly 10\% clipping.  The unsaturatd exposure taken an hour and a half later and at higher airmass had considerably less flux, 16000ADU peak, continuum of 600-1100ADU, and only 130ADU in the absorptive component of the line.  The calculated fwhm was 14.1 pixels (1.83").   

	The wavelength dependence of the psf properties shows striking changes across the line, shown in figure \ref{fig:specast-pcyg}.  Both exposures and all reductions show a similar change in centroid location, sharply changing by about 50-150mas in the blue side of the absorptive component, then rapidly dropping across the center of the absorptive component to -20mas to -80mas in the red side of the absorption.  The centroid then gradually rises over the emission to a peak of 5-15mas just before the emission line returns to continuum on the red side of the line.  The change is actually greater in magnitude for the unsaturated exposure with any reduction parameters.  There are actually six curves in the top panel of figure \ref{fig:specast-pcyg} but the difference between order widths of 41 and 47 is completely negligible.  The tilted coordinates show systematically lower deviations for both saturated and unsaturated exposures.  In essence, more flux gives less deviation presumably from better fits.  The tilted coordinates more accurately reflects the true wavelength dependence of the orders and reduces the systematic error.  
	
	The fwhm also shows strong differences between the different exposures and reduction parameters.  The unsaturated exposure with only 140ADU peak flux shows a 200mas decrease in fwhm in the absorptive component of the line in both tilted and untilted coordinates with a width of 47 pixels.  When the wings of the psf are neglected by using a width of 41, the calculated fwhm deviations are significantly smaller, near 75mas, presumably because the wings of the psf have a noise contribution.  This again shows that more flux leads to better fits and that large flux differences across a line lead to systematic effects.  The saturated exposure shows an extra 50mas increase in fwhm across the emission peak in the clipped region of the line for all reductions.

 %%%%%%%%% Figure 13 here

\begin{figure}[!h!t!b!p]
\hspace{3mm}
\includegraphics[width=0.7\linewidth, height=0.9\linewidth, angle=90]{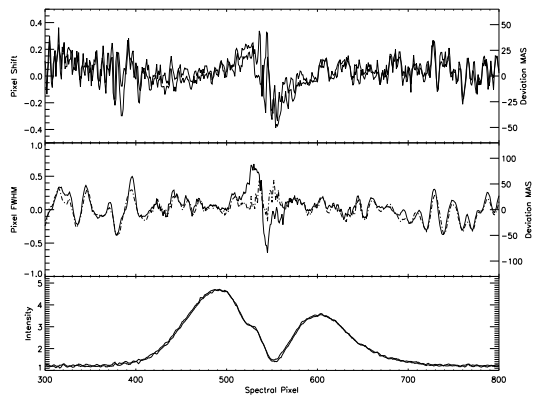}
\figcaption[specast-tori]{\label{fig:specast-tori} The wavelength dependence of the psf for T-Ori reduced with and without tilted coordinates.  The top box shows the center of a Gaussian fit to the PSF at every spectral pixel.  A systematic centroid shift effect is seen across the emission peak of line where there is extended, resolved H$\alpha$ emission.  Using a tilted coordinate grid does not change the deviation.  The middle box shows the calculated FWHM for each spectral pixel, smoothed for clarity.  The tilted coordinates actually make the fwhm variation across the emission negligable, showing that this resolved emission does *not* contribute substantially to the fwhm.  The bottom box shows the resulting normalized intensity.  }
\end{figure}  
	
\begin{figure}[!h!t!b!p]
\hspace{3mm}
\includegraphics[width=0.75\linewidth, height=0.95\linewidth, angle=90]{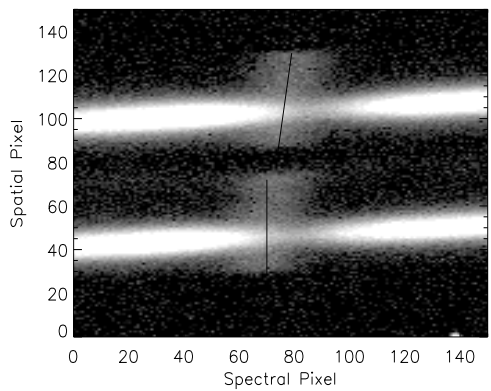}
\figcaption[tori]{\label{fig:tori} The raw exposure of T Ori showing the spatially resolved extended H$\alpha$ emission near the absorptive component of the stellar line.  The lower beam has an untilted spectral slice shown, and the upper beam has a tilted coordinate grid overplotted.  This illustrates a systematic error that may result from extended resolved nebular emission around a source causing a change in centroid and fwhm.  }
\end{figure}

	Another complication is extended emission.  A change in fwhm and centroid can be caused by significant resolved emission around an object, such as our observations of T Ori.   If there is no inspection of the raw data, this emission may be falsely reported as a binary detection.  Also - one cannot rule out the possibility of unresolved nebulosity as a significant contribution to the psf.  The observations of T Ori showed a peak signal of 600ADU with a continuum of 100-150ADU.  The fwhm was calculated as 10.1 pixels (1.3").  The wavelength dependence of the psf for two reductions, using tilted and untilted coordinates, is shown in figure \ref{fig:specast-tori}.  The centroid shift is small, roughly 40mas and is independent of the order tilt.  The fwhm changes by roughly $\pm$80mas when using untilted coordinates, but is essentially zero when using tilted coordinates.  This can arise when a vertical order geometry intersects the fully-resolved emission near the edges of the order.  When using properly tilted coordinates, the fully resolved emission becomes an added background that does not influence the Gaussian fitting.  Figure \ref{fig:tori} shows the raw exposure and the extended emission with the tilted coordinates shown as the slanted black line across the extended emission.

 %%%%%%%%% Figure 14 here

	Further investigation of these effects is beyond the scope of this paper, but the analysis package is robust and subject to the same systematic errors as other analysis packages (Bailey 1998, Baines et al. 2006).  In summary, this analysis package can reproduce results of others, specifically MWC 166.  However, other systematic effects that always influence spectroscopic data must be considered as well.  This illustrates the robustness of our reduction geometry and coordinate grid as well as the psf fitting and analysis by the optimal spectral extraction routines.

     \subsection{Polarization calculation}

  The simultaneous imaging of orthogonal polarization states also allows for a greater efficiency observing sequence and for reduction of atmospheric effects since polarization measurement can be done with a single image and both states are imaged simultaneously, with identical systematics.  Using the fractional polarizations measured as the difference between the orthogonally polarized spectra divided by the sum, q and u, we can calculate the percent polarization, P, and the position angle, PA, as follows:

\begin{equation}
q= \frac{Q}{I} = \frac{1}{2}(\frac{a-b}{a+b} - \frac{c-d}{c+d})= \frac{1}{2}(q_{0^\circ} + q_{45^\circ})
\end{equation}

\begin{equation}
P = \sqrt{q^2 + u^2}
\end{equation}

\begin{equation}
PA = \frac{1}{2}tan^{-1}\frac{q}{u}
\end{equation}

  We wish to note that many people use the ratio method defined as follows:

\begin{equation}
q= \frac{Q}{I} = ( \frac{ \sqrt{\frac{ad}{bc}}-1 }  { \sqrt{\frac{ad}{bc}}+1})
\end{equation}

  The Ukirt infrared polarimeter has yet another difference method published to calculate the polarimetry on their IRPol2 website.  
  
\begin{equation}
q= \frac{Q}{I} = (\frac{a-c-\frac{a+c}{b+d}(b-d)}{a+c+\frac{a+c}{b+d}(b+d)})
\end{equation}

  This method is a different version of the simple average method and can be reduced to a simple average where b and c switch places:
  
\begin{equation}
q= \frac{Q}{I} = \frac{1}{2}(\frac{a-c}{a+c} - \frac{b-d}{b+d})
\end{equation}

	When all three of these methods are applied to a data set, the difference between the calculated polarized spectra is entirely negligable.  Since no method was significantly different than any other, we use the easiest to implement - a simple average applied to each individual exposure set.  
	
	We wish to note that no instrumental polalarization ripple of the kind mentioned in Aitken \& Hough 2001 or seen in ISIS (Harries et al. 1996) has been seen in our data.  This ripple is thought to arise from Fabry-Perot type fringes in the wave-plate.  Though our instrument uses a standard achromatic waveplate, a thin layer of polymer sandwiched between two glass substrates, no ripple near the 0.1\% level has been seen.

	\subsection{Aligning the spectra}

  The Savart plate axis can never be perfectly aligned with the slit.  This produces a small shift in the dispersion direction between the two polarized orders. This shift was calculated by cross-correlating the ThAr spectra from the two polarized orders.  The derived shifts were 2 pixels in 2004, 12 pixels in 2005 before the Savart plate reinstallation, and 7 pixels after the 2006 remounting.  
  
\begin{figure}[!h!t!b]
\includegraphics[width=0.7\linewidth, height=1.0\linewidth, angle=90]{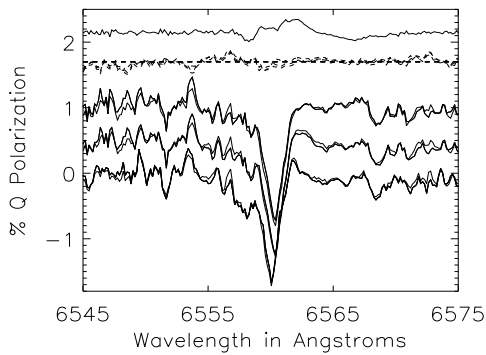}
\figcaption[q9ways]{\label{fig:q9ways} Stokes q calculated 9 different ways and the differences between them- from the bottom, no alignment, integer-pixel alignment, subpixel-alignment, and the differences between all of them and the first one (simple average, no alignment).  They're all quite similar in average shape.  The changing S/N can be seen across the trough and peak of the PCygni line.  The noise increases when any spectral alignment procedure is applied.   }
\end{figure}

  To characterize any derivative noise, there are a number of ways we can align the spectra and calculate the polarization.  The three ways are 1) no alignment 2) cross-correlation an a simple integer-pixel shift (which preserves noise properties) and 3) integer-pixel shift followed by the shift-n-scale sub-pixel alignment.  When no alignment is done, the spectra and polarization are simply calculated where the are on the detector.  In the integer-pixel-shift alignment, we do a cross-correlation on the spectra to ensure that the spectra aligned to a single pixel, as a double-check on the wavelength calibration.  Since any interpolation can effect and corrupt the noise statistics of the spectra, as shown by the shift-n-scale example in figure \ref{fig:shfscl}, subpixel alignment must minimize the derivative signal more than it introduces extra noise.  To evaluate all three alignments, the polarization analysis is done on the three different spectra using the three different alignment methods.  These nine polarization spectra are similar in their overall structure.  A sample of this is shown in figure \ref{fig:q9ways}.  The polarization signature is around 1.5\% and the differences between the different methods are very small.

 %%%%%%%%% Figure 15 here

  The spectral shift (beam wobble) measured from Th-Ar lamp spectra taken at many waveplate orientations shows that there is a one pixel shift induced over a complete rotation of the waveplate with a roughly sinusoidal shape.  The alignment of the waveplate with the savart plate causes a rotational offset between the rotation-stage zero point and the waveplate zero point of 85$^\circ$.  This means we operate our waveplate from rotation angles of 85$^\circ$ to 152.5$^\circ$ near the crest of the sine-curve.  

\begin{figure}[!h!t!b]
\includegraphics[width=0.7\linewidth, angle=90]{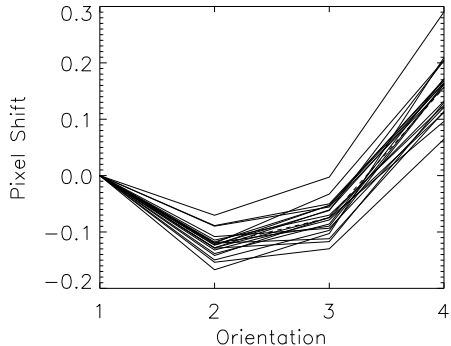}
\figcaption[wobble-exp]{\label{fig:wobble-exp} The beam wobbles (shows a small wavelength shift) as the waveplate rotates.  The wobble is calculated by applying the shift-n-scale routine independently to many lines in the Th-Ar lamp spectra.  The wobble shows a sinusoidal dependence on waveplate angle with an amplitude of 1 pixel.  The 67.5$^\circ$ region of operation is near the crest of the sine-curve and hence has an amplitude of 0.4 pixels.}
\end{figure}

  Measurements of ThAr spectra taken each night, shown in figure \ref{fig:wobble-exp}, show that the beam wobble is roughly 0.3 pixels between the four waveplate orientations.  The derivative signal caused by a 0.3 pixel shift for strong emission lines with contrast of 5-20 is roughly 1-3\%.  It will be shown later that the derivative signal subtracts away in the dual-beam polarization analysis and that a beam wobble of 0.3 pixels produces a negligible error.  This beam-wobble signal follows regions of curvature and is proportional to the initial derivative error, as it is a derivative-of-derivative error (acceleration).  We also calculated the sub-pixel shifts between the three different intensity calculation methods using the shift-n-scale routine.  We found the shifts to be a small fraction of a pixel. 
  
 %%%%%%%%% Figure 16 here

	The wavelength of each pixel for each horizontal order is assigned by a 3$^{rd}$ order polynomial fit to thorium-argon calibration lamp images.  This analysis is done independently on each order for the two orthogonal polarization states.  Typically, reduction packages, such as Libre Esprit for the ESPaDOnS spectrograph on CFHT (Donati et al. 1997) use the wavelength fits as the alignment method between the polarized orders.  We have found that the simple average without alignment method works best because the flat-field cancels in the polarization calculation, leaving the measurement independent of the pixel-to-pixel gain.

	\subsection{Rotating Frames}

  Since HiVIS uses the alt-az coud\'{e} focus, the spectropolarimeter's Q-U reference frame rotates on the sky as the telescope tracks.  There are two main rotations: one with altitude, between the tertiary mirror and m4, as the altitude changes, the other with azimuth, between m6 and the coud\'{e} pickoff mirror.  Normally, coud\'{e} spectrographs use a 3-mirror image rotator somewhere downstream (in an pupil plane) to compensate for the telescope's motion and keep the orientation of the slit on the sky fixed.  We have an image rotator, but since the rotator is 3 oblique reflections, we do not use this rotator to increase the accuracy of the polarimetric calibration of the instrument.  Changing the mirror orientations would change the polarization induced by the mirrors.  The orientation of the instrument on the sky is allowed to rotate and exposures are kept short so that the rotation is small.  
  
  The analysis software takes the date and time of the exposure and calculates the altitude and azimuth of the telescope during the exposure.  Knowing the orientation of the telescope mirrors, we can reconstruct the projection of the spectropolarimeter's Q and U axes on the sky, and rotate all measurements to a common frame.  A common convention is to set +Q to North on the sky. 
  
  The rotation angle between the 4 individual exposures in each set is calculated as Position-angle + Altitude - Azimuth and is checked to quantify the inter-set smear.  Very large rotation angles occured when an object is tracked near transit (high azimuth rates).  For measurements with small rotations, the polarization measurements are assumed to be accurate and are rotated to an absolute frame (of the first exposure) on the sky.

  For every data set, rotation in an individual exposure causes each exposure to become a mix of Stokes Q and U, reducing the efficiency of the measurement and adding some uncertainty to the Stokes parameters.  Imagine an incident +Q beam.  A rotation of the instrument of 22.5$^\circ$ during an exposure would cause the projected instrument frame to go from +Q to +U and the resulting measurement would only show 50\% polarization, when 100\% was incident. 
  
    Rotation between sets also induces systematic error.  As the instrument rotates on the sky, different amounts of time will be spent sampling the polarization states.  As an extreme example, 22.5$^\circ$ rotation of the instrument between each individual exposure would cause all images to be oriented on the sky with +Q whereas the reduction package would assume the sequence was +Q, +U, -Q, -U.  The systematic error with this high rotation rate causes complete loss of information about U while doubling the time spent measuring Q.  It also causes the benefits of beam-swaping to be lost entirely and it causes the reduction methods to fail.

\subsection{Rebin By Flux}

 %%%%%%%%% Figure 17 here

\begin{figure}[!h!t!b]
\includegraphics[width=0.6\linewidth,height=1.0\linewidth, angle=90]{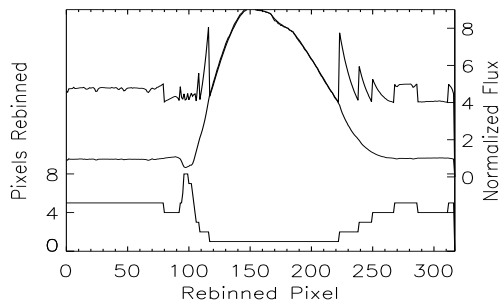}
\figcaption[rb]{\label{fig:rb} The rebin-by-flux routine adaptively averages pixels to equalize the S/N per spectral pixel.  The number of physical pixels averaged per rebinned spectral pixel in the adaptive rebinning with a threshold of 4 is the bottom curve.  In the P-Cygni absorption trough, 8 pixels are binned, but in the emission peak, there is no binning. The middle curve is the rebinned spectra which shows a highly compressed P-Cygni trough because of the high binning.  The top curve is the flux per rebinned spectral pixel, with a minimum of 4 (the threshold) and a maximum of the emission peak. }
\end{figure}

  Since the expected spectropolarimetric signals are subtle differences in the polarized line profile shapes of less than one percent, and the few reported detections show polarization of at most 1\%, the lines must be calculated with at least S/N$ >$ 300 in all spectral-resolution-elements.  
  
    Most young emission-line stars have large line/continuum ratio's (5-10).  Some, such as AB Auriga, have PCygni profiles with emission-peak to absorption-trough ratio's of over 20.  The star P-Cygni itself has a ratio of nearly 100.  Even when saturating the peak of the emission line on the detector, a S/N of only 100 or so was achieved in the absorption trough of AB Auriga.  A flux-dependent averaging procedure has been developed to bin pixels spectrally by flux, to ensure that the S/N per spectral-pixel is roughly constant over the line at the expense of regular wavelength coverage.   

 %%%%%%%%% Figure 18 here

\begin{figure}[!h!t!b]
\hspace{2mm}
\includegraphics[width=0.8\linewidth, height=1.0\linewidth, angle=90]{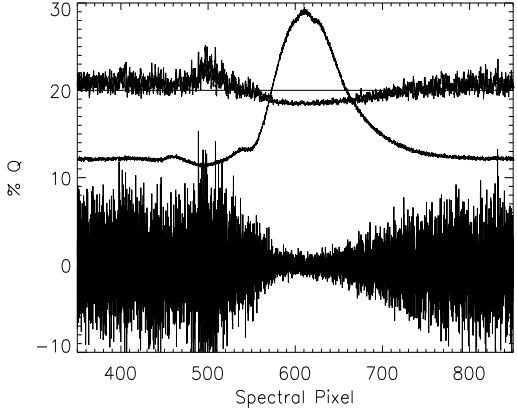}
\figcaption[qnoise]{\label{fig:qnoise} The systematic error in a simulated data set caused only by varying noise.  Random noise proportional to $\sqrt{N}$ was added to eight identical copies of the H$\alpha$ line of AB Aurigae, plotted above.  The resulting Stokes parameters were calculated and show a noise of roughly 5\% at continuum and 2\% in the emission line peak.. The raw Stokes q and u are the bottom curves, centered about zero, with the noise obviously varying across the line profile.  The spectropolarimetric signal, $\sqrt{q^2+u^2}$, is the top curve and it varies by over 2\% across the line profile.  The variation is caused only by the varying noise, rising with higher noise.    }
\end{figure}

  Another systematic error occurs when you calculate the degree of polarization for a spectrum, a squared quantity, with noise amplitudes that vary across the line.  Since the average absolute values of noisy data will be higher than clean data, and there will be significant noise variations across strong emission lines, varying noise can mascarade as a systematic spectropolarimetric error.  Figure \ref{fig:qnoise} shows a simulated polarization measurement where flux-dependent noise was added to eight copies of a single line-profile.  The noise varied from roughly 5\% at continuum to roughly 2\% in the center of the emission line.  The resulting Stokes q and u parameters were centered about zero, but with a varying noise amplitude.  When these measurements are converted to a degree of polarization via $\sqrt{q^2+u^2}$, this varying noise shows up as a systematic error in the degree of polarization.

 %%%%%%%%% Figure 19 here

\begin{figure}[!h!t!b]
\hspace{1mm}
\includegraphics[width=0.8\linewidth, height=1.0\linewidth, angle=90]{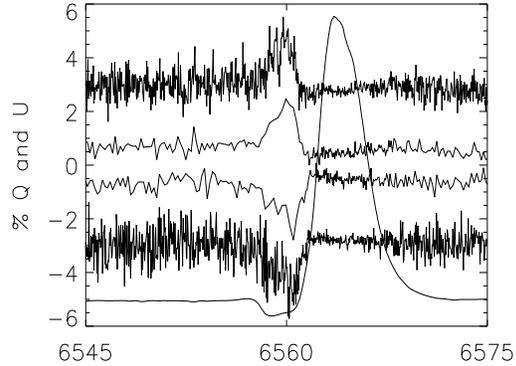}
\figcaption[rbqu]{\label{fig:rbqu} An example of the rebin-by-flux routines applied to a data set.  The noisy curves show Stokes q (top) and Stokes u (bottom) with a uniform spectral resolution, but an uneven S/N.  The smoother curves near the middle have a more uniform noise per pixel and the wavelength dependence of the polarization is much more clearly seen.  The average H$\alpha$ line is overplotted. }
\end{figure}

 To rebin-by-flux, the average line profile for each data set (after alignment) is calculated.  This provides the average flux and shape of all 8 lines.  A user-defined threshold, typically 1-5, sets the minimum flux in each rebinned spectral pixel.  The number of physical pixels to rebin at each wavelength is calculated and each individual spectrum is then rebinned accordingly.  A rebinned wavelength array for each set is also calculated since the number of pixels binned in the spectral direction is different for each star and data-set.  The polarization analysis is then performed on these rebinned data sets as specified above and the pixel-to-pixel variation in the noise is greatly attenuated.  Figure \ref{fig:rbqu} shows a comparision of raw and rebinned polarization spectra.  The rebinned spectra only show a few resolution elements over the absorption trough with many covering the emission peak.  The polarization signature in the absorptive part of the line stands out very clearly.

\section{Systematics - Alignment Error}

 The calculated polarization is sensitive to systematic errors in the alignment of the polarized spectra.  There are two main alignments of the spectra - between the polarized spectra in each individual image, and between the specra from each individual image. 

 The reduction code automatically makes derivative-error simulations for every data set to show the form of the systematic errors caused by misalignments.   The misalignment between polarization states causes a derivative signal in each fractional polarization measurement  $\frac{(a-b)}{(a+b)}$ that can be at the few \% level.  As long as the shift between polarization states is constant in each image, then the derivative signal is subtracted away when the second polarization measurement, $\frac{(c-d)}{(c+d)}$, is subtracted.  The polarization state shift between images is very stable.  There is never more than a 0.05 pixel shift difference between the calculated wavelength solutions of each polarized order as determined by ThAr line fits.  
 
 %%%%%%%%% Figure 20 here

\begin{figure}[!h!t!b]
\hspace{2mm}
\includegraphics[width=0.7\linewidth, height=1.0\linewidth, angle=90]{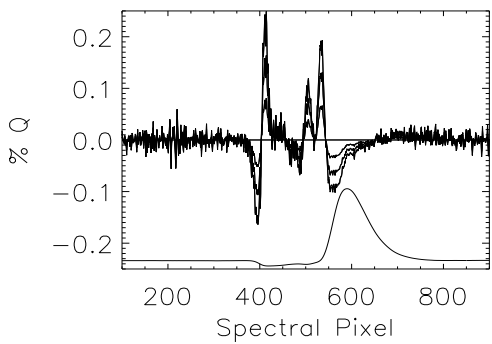}
\figcaption[derivsim]{\label{fig:derivsim} A beam-wobble polarization simulation - modeling the systematic error in polarization caused by a spectral shift of the spectra induced by beam wobble or optical misalignments.  As the waveplate rotates, the spectra of one orientation shift by a small fraction of a pixel to those of another orientation.  This simulation greatly exaggerates the shift to show what a strong systematic error would look like.  The curves show the simulated Stokes Q spectra for beam wobble amplitudes of 0, 0.5, 1.0, and 1.5 pixels.  The largest shift shows a peak-to-peak amplitude of 0.4\%.  The systematic effect follows the curvature since beam wobble is a derivative of derivative. }
\end{figure}

  There is also a second order effect - beam wobble - that can cause a systematic polarization error.  If the spectra are misaligned between images, from the rotating waveplate or guiding errors in un-tilted coordinate systems, then a derivative-of-derivative effect can show up when the two polarization measurements are differenced.  We simulate this effect in figure \ref{fig:derivsim} by applying a orientation-dependent wavelength shift between the exposures that increases with waveplate angle.  This simulates any beam-wobble introduced by misalignment of the waveplate.  The induced error traces the curvature of the emission line, with polarization errors showing up where curvature is greatest.  Using ThAr spectra, we measured this wavelength shift to be $\sim$0.1 pixels, depending on waveplate orientation, shown in \ref{fig:wobble-exp}.  This small wobble has a very small influence on the polarization, as it takes a misalignment of several pixels to produce a noticable acceleration signal.  
  
  The derivative simulations in this code show that the simple derivative error subtracts away completely if the polarized-order-separation is constant (which it is to 0.05 pixels).  The code also shows that it takes a wobble of at least 2 full pixels between each orientation, for a total of a 6-pixels misalignment to produce a 1\% systematic error in a Stokes parameter for strong lines.  Using our shift-n-scale routine, as well as the ThAr measurements, we have measured this wobble to be around 0.1 pixels and thus present only well below the 0.1\% level proportional to the curvature (derivative of derivative).

%============================================================================================================================

\section{Polarization Calibration - The AEOS Model}

	The next step in the instrument development was calibrating the instrumental polarization response.  Absolute polarization at all wavelengths requires very careful calibration of the telescope.  Internal optics can induce or reduce polarization (U,Q$\rightarrow$I or I$\rightarrow$Q,U) in the beam, cause cross-talk between polarization states (QU$\rightarrow$V, V$\rightarrow$QU), and also rotate the plane of polarization (Q$\rightarrow$U, U$\rightarrow$Q).  For an instrument at the coud\'{e} focus of an alt-az telescope, all of these effects are functions of wavelength, altitude and azimuth.  For AEOS, there are five 45$^\circ$ reflections before the coud\'e room, two of which change their relative orientation.  This is particularly severe, and only a few put a spectropolarimeter after so many mirrors.  The calibration is difficult, but some have tried this for many different telescope designs (cf. S\'{a}nchez Almeida et al. 1991, Kuhn et al, 1994, Giro et al. 2003) 
	
	Since there are no spectropolarimetric standards one usually observes polarimetric standards that have a precise value averaged over some bandpass such as Hsu \& Breger 1982, Gil-Hutton \& Benavidez 2003, Schmidt \& Elston 1992, or Fossati et al. 2007.  The calibration and creation of a telescope model is done by measuring unpolarized standard stars and polarized sources at many pointings to calibrate the effects of the moving mirrors.  
	
	In general, the telescope will have different responses to linearly polarized light since the absorption coefficients of the optical components vary with wavelength and the differential absorption is a function of the angle of the incident polarization with respect to the mirror/component axes.  We used Zeemax models of the telescope to compute the assumed polarization properties of the coud\'{e} focus given various optical constants for aluminum.  These show the expected altitude-azimuth dependence of the mueller-matrix terms, though the amplitudes of the computed effects are significantly different than measured.  While our models can calculate the full Mueller matrix as functions of altitude, azimuth, and wavelength, we cannot measure all of these terms.  We can, however, measure certain parts of the Mueller matrix given proper sources (polarized standards, unpolarized standards, and scattered sunlight).  

	Unpolarized standard stars are the first tool we used to calibrate the telescope.  Observations of these standards at as many pointings as possible allows one to construct a model of the telescopes response to unpolarized light and to quantify the telescope-induced polarization (but not the depolarization, cross-talk, or plane rotation which requires a linearly polarized standard).

	Polarized sources were the last calibration source we used.  Using a polarized source allows one to measure the depolarization and rotation of the plane of polarization, all of which are, in principle, orientation dependent.  Polarized standard stars are polarized in their continuum light, but the constant value quoted in standard references is an average polarization in a relatively line-free, wide bandpass image.  The degree of polarization is usually not more than a few percent making instrument calibration difficult with these sources.  Another calibration method which we will explore below, is to use twilight (scattered sunlight) as a highly polarized source with a fairly well known degree and orientation of polarization.  The polarization is certainly not constant and varies with atmospheric properties, time, and pointing, but it is a very bright and highly polarized source that has given us valuable insight into the polarization properties of the mirrors.

	\subsection{Flat Field Polarization: Fixed Optics Polarization}

\begin{figure}[!h,!t,!b]
\hspace{-2mm}
\includegraphics[ width=0.7\linewidth, height=1.05\linewidth, angle=90]{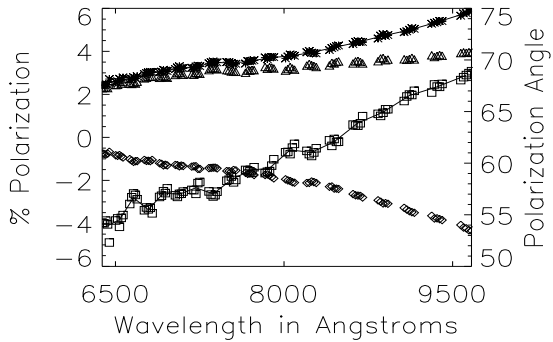}
\caption[flatpol]{\label{fig:flatpol}
Flat field polarization analysis, averaged 200:1 for clarity. Stokes q is shown as diamonds , u is shown as triangles, and the degree of polarization is the top curve, shown as + symbols.  The squares represent the position-angle of polarization (PA) with the y-axis scale on the right.  The polarization (top curve) begins at 650nm being dominated by u (triangles), but begins to rise as -q (diamons) gets larger.  The PA rotates by about 25$^\circ$, going from nearly 55$^\circ$ or mostly +u to being nearly 70$^\circ$ or a mix of -q and +u near (-3,4). }
\end{figure}

	All of the moving mirrors in the AEOS telescope are located upstream of the polarizing optics (analyzer).  As an illustration of the polarization induced by the reflections between the halogen flat field lamp ($T_{eff} \sim 2900K$) and the spectropolarimeter, figure \ref{fig:flatpol} shows the polarization reduction applied to a set of flat field frames with the 4 waveplate orientations taken in 2004.   The flat field lamp is located just after the coud\'{e} room entrance port, immediately after the coud\'{e} pickoff mirror (m7).  The lamp is reflected off a diffuser screen and is assumed to be unpolarized.

 %%%%%%%%% Figure 21 here

	The flat fields show a significant polarization which changes with wavelength.  At 6500$\AA$, the induced polarization is dominated by the +U term, but the U and -Q terms become equal at 90. The position angle varies almost linearly across all wavelengths while the degree of polarization rises from 2\% at 650nm to 6\% at 950nm.  Since the remounting of the optics and a movement of the image-rotator, the values of the flat field polarization have changed, but this illustrates the magnitude and wavelength dependence of polarization induced by the 11 fixed reflections in the optics room. 

	We also have tested the polarization of the common-fore-optics.  A linear polarizer was mounted (with a mask) at various points along the optical path, from the calibration stage just after the coud\'{e} port to just in front of the slit.  A linear polarizer was measured to have 98.2\% polarization at the slit, 98.3\% polarization after the k-cell, 97.4\% before the k-cell, and 97.7\% after the flat field screen.  This shows that the linear polarization does not vary much between the beginning of the spectrograph optics and the slit.  Another polarizer mounted at the slit was measured to have 98.4\%.  This polarizer was put at the calibration-stage and was measured to have 98.1\% 97.7\% 97.8\% and 98.0\% polarization for angles -03.2, 44.0, 87.5, and 134.2 degrees in the slit's QU frame when manually rotated to 0, 45, 90, and 135 degrees.  The measured angles match completely within the manual uncertainty.  This shows that the degree of linear polarization is preserved through the spectrograph optics.
	
	The flat field also shows a difference in beam intensities (vignetting or beam transmission difference), in addition to it's intrinsic polarization.  The polarization in a single frame, computed as $\frac{a-b}{a+b}$ shows +q at  -4.55\% -q at 1.36\% +u at  -3.63\% and -u at 0.45\%.  These values are not symmetric about zero, as would be the case if there was only intrinsic polarization.  To make the polarization values symmetric about zero, a constant difference of 1.59\% between the two beams is applied.  This was calculated as 1.588 and 1.591 for q and u respectively as the null polarization spectra (sometimes called the check spectra: +q + -q).  Restated, this means the $\pm$q $\pm$u spectra are symmetric about 1.59\%, ie that the rotation of the waveplate swaps polarization around a constant difference in beam intensity of 1.59\%.  Subtracting the 1.59\% difference in beam intensities from all single-frame polarizations $\frac{a-b}{a+b}$ makes these polarization spectra symmetric about 0\% reflecting the true polarization of the beams.  
	
	We note that the old flat field polarization measurements showed a polarization of 2.78\% at a PA of 55.0$^\circ$ (-0.93\% 2.57\% qu) for 6567$\AA$.  This is not what we measured with the new setup.  This is expected since the image rotator was moved, the flat-field lenses were installed, and the new dekkar is slightly smaller than the old one, changing the illumination pattern.  The new flat field polarization is just as repeatable and robust as under the old configuration.

 %%%%%%%%% Table 1 here

\begin{table}[!h,!t,!b]
\begin{center}
\caption{Flat field polarizations \label{flatpol}}
\begin{tabular}{lllll}
\hline
{\bf Name:}          & {\bf +q}           & {\bf -q}           & {\bf +u}            & {\bf -u}     \\
Pol \%:                 & -4.546           & 1.364             &   -3.629           & 0.452     \\
Pol-1.59\%:        & -2.956            & 2.954             &  -2.039             & 2.042      \\
\hline
\end{tabular}
\end{center}
\end{table}

	The calculated flat field polarization after background correction is -2.95\% for q and -2.0\% for u calculated with the average method as 0.5*( +q - -q ) giving a degree of polarization of 3.59\% at an angle of 287.3 degrees.  We performed this test on a sequence of 40 flat field frames and each individual exposure set gives the same answer to 0.01\%
	
	We did a test of the dichroic for the LWIS mounted inbetween m4 and m5.  The dichroic is roughly 12" square and causes an estimated 12\% reduction in throughput.  We measured the twilight polarization without, with, and again without the dichroic mounted.  The insertion and removal takes less than one minute so the twilight polarization and PA can be assumed constant throughout the $\sim$5 minute observation.  The measured polarization values were nearly identical with and without the dichroic.

\subsection{Unpolarized Standard Stars - Telescope Mirror Induced Polarization}
  
 %%%%%%%%% Figure 22 here

\begin{figure}[!h,!t,!b]
\includegraphics[ width=0.8\linewidth, angle=90]{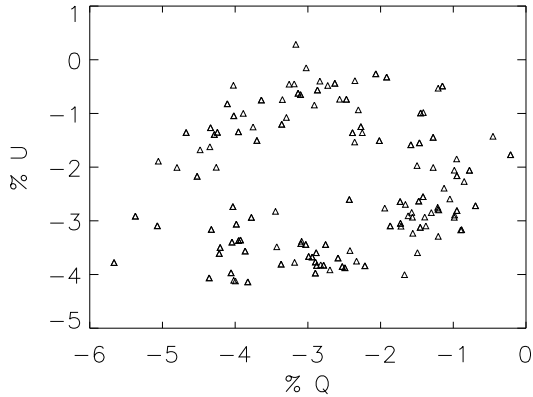}
\caption[unpol-qu]{\label{fig:unpol-qu}
The measured polarization of unpolarized standard stars for the H$\alpha$ region plotted in the QU plane. The average value is -2.74\% q  -2.38\% u or an average of 3.8\% at 295$^\circ$, which is essentially the flat-field polarization.  There was no obvious correlation with these points and either altitude or azimuth. }   
\end{figure}

 %%%%%%%%% Figure 23 here

\begin{figure}[htb]
\includegraphics[width=0.7\linewidth, height=1.0\linewidth, angle=90]{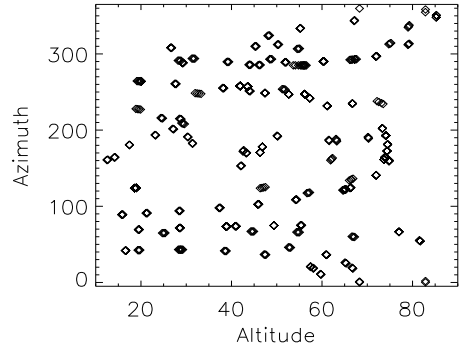}
\caption[alaz-unpol]{\label{fig:alaz-unpol}
The altitude-azimuth coverage for all the unpolarized standard stars used in making the telescope-induced polarization maps.}
\end{figure}

	Observations of unpolarized standard stars allow us to measure the variation of the telescope-induced polarization as a function of pointing.  Since the incident light is unpolarized, any observed polarization signature directly traces the polarization induced by the telescope - the IQ IU mueller matrix terms.  Many unpolarized standards have been observed in November 2004, June \& July 2005, Sept 2006 - Jan 2007, and June 2007 totaling 224 data sets.  Since the camera was changed in September 2006 and a better wavelength solution was obtained, only the 126 observations after September 2006 were used.  The calculated degree of polarization for each unpolarized standard observed is averaged to a single measurement over the entire H$\alpha$ region.  Note that all our unpolarized standards showed no significant line polarization, and had a statistical error of less that 0.05\%, typically 0.02\%.  These polarization measurements are shown on the q-u plane in figure \ref{fig:unpol-qu}.  The measurements are centered around the flat-field value of -2.95\%q -2.04\%u.  When adjusted for the flat field polarization, they show a range of polarizations from 0.5\% to 3.0\%

 %%%%%%%%% Figure 24 here
 
\begin{figure}[!h,!t,!b]
\includegraphics[ width=0.8\linewidth, angle=90]{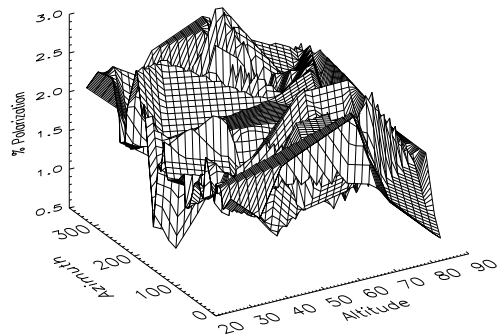}
\caption[unsurf0]{\label{fig:unsurf0}
The measured polarization of unpolarized standard stars for the H$\alpha$ region in the alt-az plane. The range of polarization is 0.5\% to 3\% for 130 independent observations. }
\end{figure}

%%%%%%%%% Figure 25 here

\begin{figure}[!h,!t,!b]
\includegraphics[ width=0.8\linewidth, angle=90]{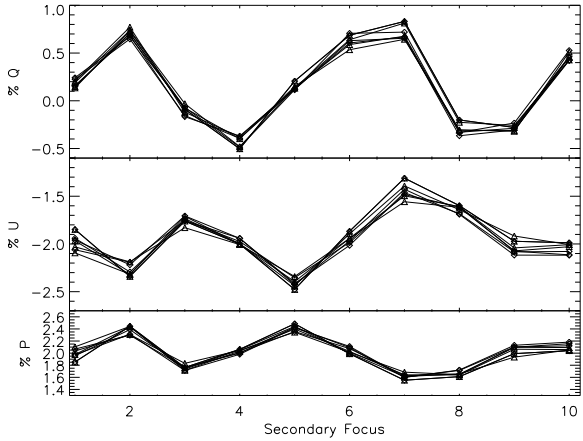}
\caption[unpol-foc]{\label{fig:unpol-foc}
The measured polarization of an unpolarized standard star - HR8430 - for the H$\alpha$ region as a function of the secondary mirror focus. There is a strong dependence on focus value.  The best focus value was 5 and the width of the psf more than doubled for focus values of 1 and 10. }   
\end{figure}
	
	 The observations can be plotted on an altitude-azimuth plane to make an all-sky induced-polarization map.  The altitude-azimuth coverage, shown in figure \ref{fig:alaz-unpol}, was fairly uniform from altitudes of 20 to 80 degrees.  There is a slight gap low in the North that comes from a lack of unpolarized stars in this region - the highest declination star we are aware of is at +50 (HR7469).  These observations are interpolated to a regular alt-az grid.  Figure \ref{fig:unsurf0} shows the all-sky map of the telescope-induced polarization.  This surface does not show a significant alt-az structure.  To understand the apparent variability in the map, we did a test of an unpolarized standard star, HR8430, as a function of the secondary mirror focus.  There is a significant variability, or order 1\% as the focus changes shown in figure \ref{fig:unpol-foc}.  This magnitude of variability with secondary focus, combined with the induced-polarization map suggest that the telescope induced polarization map has a very small alt-az dependence, of the same order as the focus dependence.

\subsection{Scattered Sunlight - A Linearly Polarized Source for Calibration}

 %%%%%%%%% Figure 26 here

\begin{figure}[!h,!t,!b]
\includegraphics[ width=0.95\linewidth, angle=90]{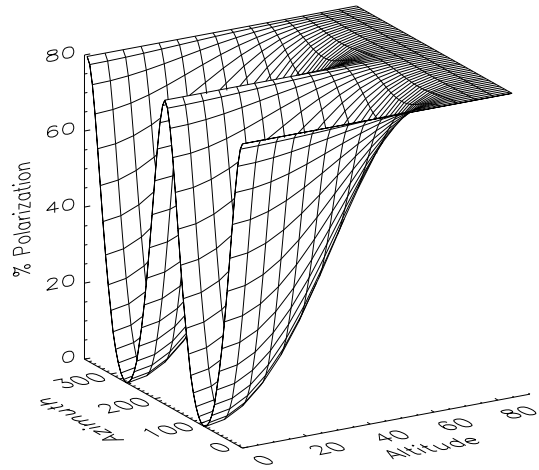}
\caption[skyim]{\label{fig:skyim}
The Rayleigh model - the degree of polarization of scattered sunlight with the sun setting in the west projected onto the alt-azi plane.  The polarization is highest where the image is white.  The maximum polarization was set to 80\% }   
\end{figure}

	Scattered sunlight is a strong linearly polarized source and is reasonably described by a simple single-scattering Rayleigh model (Coulson 1988).  Singly-scattered light is polarized orthogonal to the scattering plane with the degree of polarization proportional to $sin^2 \theta$.  The twilight polarization is highest at a scattering angle of 90$^\circ$ which reaches a maximum 90$^\circ$ from the sun on the North-Zenith-South great circle when the sun is in the West.  The degree of polarization in a Rayleigh atmosphere can be simply described with this equation:
	
\begin{equation}
d=d_{max} \frac{\sin{g}^2}{1+\cos{g}^2}
\end{equation}

\begin{equation}
\cos{g}=\sin{t_z}\sin{t}\cos{p}+\cos{t_z}\cos{t}
\end{equation}

Where g is the angular distance between the pointing and the sun, $t_z$ is the solar elevation, t is the pointing-to-zenith angle, and p is the pointing-to-solar meridian distance.  Contained the variable g is the altitude-azimuth dependence for a given location of the sun.  At sunset, $t_z\sim$0, simplifying the model.  Our observations were done with the sun lower than about 15$^\circ$, typically from 45 minutes before sunset to just after sunset.  Using this simplification, we can rewrite $\cos{g}$ as $\sin{t}\sin{p}$ and simplify the degree of polarization as:

\begin{equation}
d=d_{max} \frac{\sin{g}^2}{1+(\sin{t}\sin{p})^2}
\end{equation}

This gives us a simple square-sinuosoid in angular distance from the sun (g) modulated by the pointing-to-zenith and meridian distances.  The North-Zenith-South meridian is 90$^\circ$ from the sun and represents the maximum polarization $d_{max}$.  The plane of polarization would always be parallel to the altitude axis (vertical).  In the absence of optical depth effects (which depolarize the light near the horizon), the polarization would always be $d_{max}$ on this meridian.  On the meridian from the East-Zenith-West, the degree of polarization would vary as sin$^2$, being zero in the East and West and $d_{max}$ at the zenith.  The plane of polarization would always be perpindicular to the EZW meridian and aligned with the azimuth axis of the telescope (horizontal).  Figure \ref{fig:skyim} shows this simple Rayleigh model for the twilight polarization projected onto the altitude-azimuth plane.
	
	At our high-elevation (10000ft) the dry, clear atmospheric conditions are particularly good for using a Rayleigh model because of the low optical-depth and high degree of polarization.  Near sunset at normal telescope pointings, scattered sunlight has a known position angle and a degree of polarization of typically 30-80\%.  Observing scattered sunlight at many different altitudes and azimuths allows us to check the instrument's response to linearly polarized light as a function of mirror angles.  
	
 %%%%%%%%% Figure 27 here

\begin{figure}[!h,!t,!b]
\includegraphics[ width=.8\linewidth, angle=90]{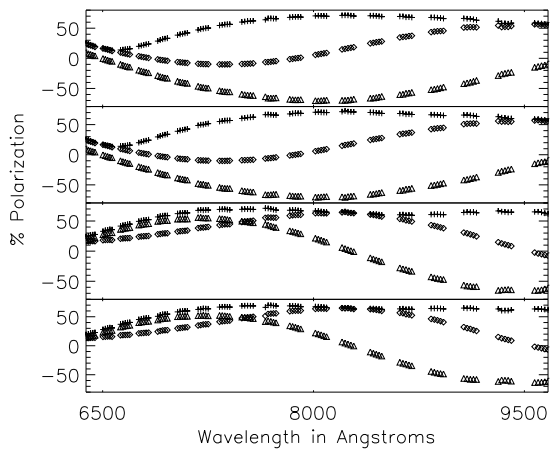}
\caption[sky75]{\label{fig:sky75}
Scattered sunlight polarization at an elevation of 75$^\circ$ for azimuths of NSEW from top to bottom.  Polarization of scattered sunlight is maximum at 90$^\circ$ scattering angle and at sunset, this is the arc from North to South through the Zenith.  The polarization spectra have been averaged 200:1 for ease of plotting.  The symbols are: q=$\Diamond$,  u=$\triangle$,  P=$\sqrt{Q^2+U^2}$=+.  P is the top curve with Stokes q and u are the more sinuosoidal curves with a strong rotation of the plane of polarization ($\frac{1}{2}tan^{-1}\frac{q}{u}$).}
\end{figure}

	This scattered light is not a perfect source however.  The degree of polarization is a strong function of the scattering properties of the air.  In particular, aerosol content (salt, dust, water, etc), optical depth (multiple scattering changing with altitude and horizon distance) and reflections from the Earth's surface (land and ocean) can significantly change the observed linear polarization (Lee 1998, Liu \& Voss 1997, Suhai  \& Horv\'{a}th 2004, Cronin et al. 2005).  Our site is surrounded by a reflecting ocean surface, and it is unknown what effect this has on the twilight polarization at our site.  Since the sun was up during most of our observations, the polarized light reflected off the ocean surface has potential to complicate our measurements.  However, we expect this effect to be small because the maximum polarization at zenith obsered over many nights was relatively constant, being $\sim 75\%$, while the cloud covering the ocean below changed significantly.  The broad trends in the degree of polarization we measured are in agreement with the singly-scattered sky polarization model, but the linear polarization observed by many researchers has been shown to be functions of time, atmospheric opacity, aerosol size distributions, and composition, all of which vary significantly from day to day (Coulson 1988, Liu \& Voss 1997, Suhai \& Horv\'{a}th 2004).  We have no accurate, quantitative simultaneous observations of the twilight polarization to compare with our observations, but there are still many useful things to learn from measurements of a highly polarized source. 

 %%%%%%%%% Figure 28 here

\begin{figure}[!h,!t,!b]
\hspace{2mm}
\includegraphics[ width=0.7\linewidth, angle=90]{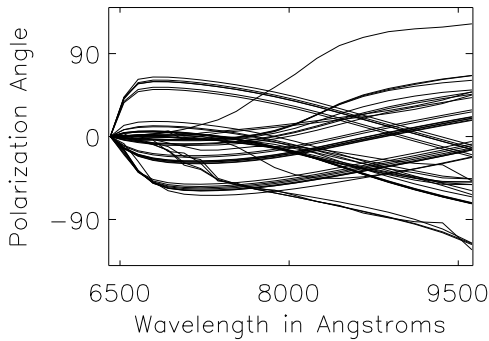}
\caption[sky_theta]{\label{fig:skytheta}
The calculated angle of polarization ($\frac{1}{2} tan^{-1}\frac{Q}{U}$) for scattered sunlight at altitudes [30,50,65,75] and azimuths [N,E,S,W] with the sun .  The 52 observations were taken June 26th to July 7th 2005 with somewhat irregular sampling on this alt-az grid.}
\end{figure}

	We previously reported observations of twilight at many different pointings from June 26th to July 7th 2005 in Harrington et al. 2006.  We obtained 52 twilight polarization measurements (416 spectra) on an altitudes-azimuth grid of [30$^\circ$, 50$^\circ$, 65$^\circ$, 75$^\circ$, 90$^\circ$] and [0$^\circ$, 90$^\circ$, 180$^\circ$, 270$^\circ$] covering 6500$\AA$ to 9500$\AA$ with the old science camera.  Due to time constraints and a much longer read-time, we were unable to obtain more than 5 measurements on any one evening, and there is significant variability at single pointings of over 5\% night-to-night.  We then compared these measurements to expected sky polarizations (Coulson 1988, Horv\'{a}th et al. 2002, Pomozi et al. 2001).  
	
	These spectra showed varied wavelength dependences of polarization and position-angle.  A sample of polarization spectra in the north, south, east, and west at 75$^\circ$ elevation are shown in figure \ref{fig:sky75}.  They have been binned to 200 times lower spectral resolution, five points per spectral order, for ease of plotting.  The figure shows the north-south and east-west pairs are almost identical even though the north and west spectra were taken one night after the south and east spectra.  Most of the 52 spectra showed stronger polarization changes in degree and angle at shorter wavelengths with the degree of polarization becoming more constant at longer wavelengths.  However, the rotation of the plane of polarization was a very strong function of wavelength and pointing, with sometimes 90$^\circ$ of rotation from 6500$\AA$ 9500$\AA$).  The strong rotation of the plane of polarization for all 2005 twilight measurements is shown in figure \ref{fig:skytheta}.  These curves do not sample the alt-az grid evenly, but they all show strong, relatively smooth rotation with wavelength.  Some show a very strong dependence between 6500$\AA$ and 7000$\AA$.  Even if the degree of polarization becomes more constant, the rotation of the plane of polarization continues with wavelength.  It is hard to explain this as anything but the telescope.   
		
	We then did much more extensive twilight observations with the new hardware in June 20th-22nd 2007.  The AEOS dome walls were raised, but the aperture was open, allowing us to observe only altitudes above 60$^\circ$.  We obtained 109 polarization measurements over the evenings of June 20th, 21st, and 22nd, typically spending an hour to completely map the sky.  On all three nights we covered at least one complete map and we always had multiple overlapping observations.  Since the sun's azimuth at sunset was 296$^\circ$ during these three nights, we aligned our azimuth coverage to reflect this as our west zero-point.  The observations spanned 4:46-5:19UT on 6-20, 4:11 to 5:01UT on 6-21 and 4:23 to 5:01UT on 6-22.  Over this time frame, the sun's altitude dropped by about 10$^\circ$ and the azimuth changed by 5$^\circ$.  The night of the 20th, we obtained one complete grid at altitudes of 60,75, and 89 degrees elevation at azimuths of 26, 116, 206, and 296 degrees, the equivalent of NESW in the simple rayleigh model.  On the subsequent two nights, we extended our azimuth coverage to include azimuths 71, 161, 251, and 341, or the secondary coordinates (NE, SW, etc).  We obtained nearly three full patterns on the 21st and nearly two full patterns on the 22nd.  
	
 %%%%%%%%% Figure 29 here

\begin{figure}[!h,!t,!b]
\includegraphics[ width=0.8\linewidth, angle=90]{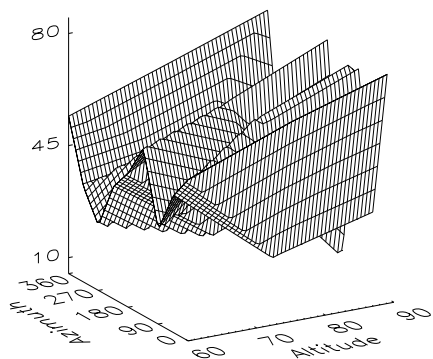}
\caption[sky-ps]{\label{fig:sky-ps}
The average measured polarization of scattered sunlight as a function of altitude and azimuth during sunset for all observations on the 20th-22nd.  The twilight polarization is assumed to follow a simple rayleigh-scattering model with an azimuth-independent maximum at zenith.  The saw-tooth form at high altitudes is entirely the telescope. }   
\end{figure}

	As with the unpolarized standards, the scattered sunlight polarization measurements can be plotted on the altitude-azimuth plane.  Figure \ref{fig:sky-ps} shows the measured degree of polarization projected on the sky for the June 2007 measurements around the H$\alpha$ line.  There is a striking 4-ridge structure in the degree of polarization that immediately stands out at the zenith.  The oscillations are strongest at high altitude and decrease in amplitude as the altitude falls, becoming double peaked by an altitude of 60$^\circ$.  The position-angle also shows strong variation with pointing.  

 %%%%%%%%% Figure 30 here

\begin{figure}[!h,!t,!b]
\includegraphics[ width=0.8\linewidth, angle=90]{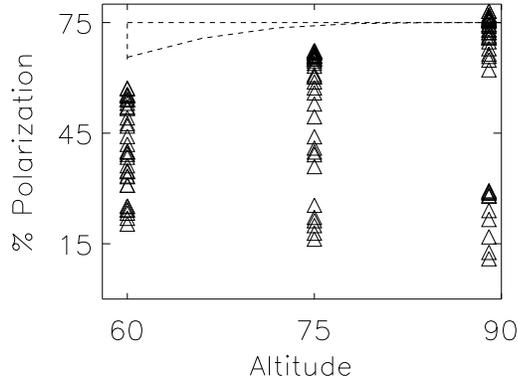}
\caption[sky-pva]{\label{fig:sky-pva}
The measured polarization of scattered sunlight as a function of altitude during sunset.  The polarization should rise with altitude, but one rising trend (along cardinal NESW azimuths) and one falling trend (along intermediate azimuths SW, SE, NW, NE) is distinctly seen. The dashed line represents the region all the points would occupy for a simple Rayleigh model with a maximum degree of polarization of 75\%.}   
\end{figure}

	Figure \ref{fig:sky-pva} shows all the measurements plotted versus altitude.  The polarization at the zenith is at its maximum in the simple rayleigh model and should not vary at all with changing azimuth.  The spread in the values at each azimuth shows the effect of the telescope on polarized light.  On all three nights there were observations at a pointing repeated an hour after each other.  The change in polarization in that hour was as much as 10\% (rising by 10\% at zenith as the sun set).  However, the change in polarization from the telescope was roughly 5 times this amount in azimuth.  In an ideal setting, the points would all converge to a maximum polarization ($\sim$75\%) at the zenith while maintaining a spread from low to high at lower altitudes.  The dashed line in the figure shows the simple Rayleigh model range for a maximum polarization of 75\%.  This is obviously not the case, and two things stand out. The first is that there are no points at maximum polarization at lower altitudes.  In the simple rayleigh model, observations to the north and south should be just as strongly polarized as the zenith.  The second is that the polarizations measured at the zenith range from 10\% to 80\%.  Since the measurements at the zenith represent a 2$^\circ$ patch of sky, and the time variability is of order 10\%, the telescope must cause an effective depolarization from 75\% to 15\% for some pointings.

 %%%%%%%%% Figure 31 here

\begin{figure}[!h,!t,!b]
\includegraphics[ width=0.8\linewidth, angle=90]{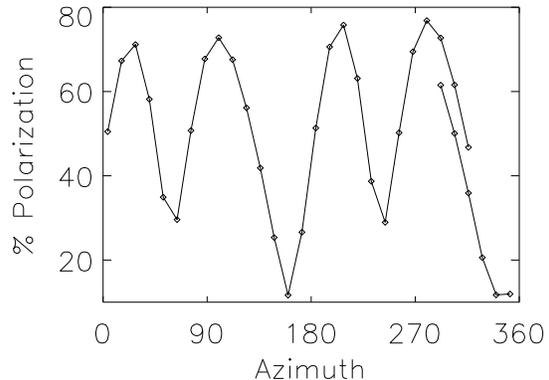}
\caption[sky89]{\label{fig:sky89}
The measured polarization of scattered sunlight at the zenith during sunset as the telescope rotates in azimuth.  The first measurement is at 292$^\circ$ at 4:16UT, wrapping around and finishing at 316$^\circ$ at 4:56UT.  The polarization should be constant with all azimuths, but a sinosoidal variation is seen.  This signature is caused by the crossing and uncrossing of the m6-coud\'{e} pickoff mirror pairs.  The minima at 160$^\circ$ and 340$^\circ$ are around 10\% whereas the minima at 70$^\circ$ and 250$^\circ$ are around 30\%.  The maxima are always 65-75\% with a systematic, nearly linear increase with time. }   
\end{figure}

\begin{figure}[!h,!t,!b]
\includegraphics[ width=0.8\linewidth, angle=90]{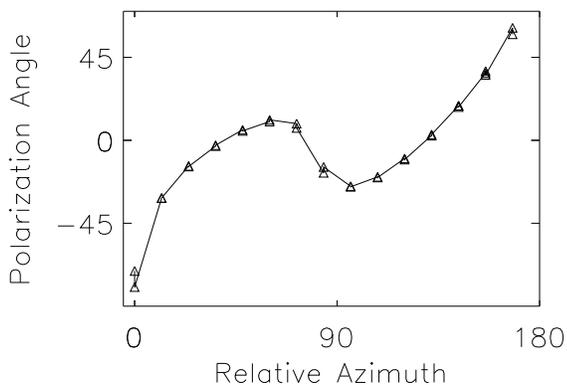}
\caption[sky89ang]{\label{fig:sky89ang}
The measured position-angle of polarization of scattered sunlight at the zenith during sunset in the instrument frame as the telescope rotates in azimuth.  Since there is a 180$^\circ$ ambiguity in the Stokes parameters, the data have been folded by 180$^\circ$ with the measured angles matching almost perfectly.  The polarization should be rotate linearly with azimuth, but an extra variation of up to 30$^\circ$is seen.  This signature is caused by the crossing and uncrossing of the m6-coud\'{e} pickoff mirror pairs.  The relative azimuth is set to 160$^\circ$ to match the minimum polarization seen in the twilight measurements. }   
\end{figure}

	In order to further examine these surfaces, we performed more twilight measurements at the zenith on July 27th and 28th 2007.  Both nights were relatively clear but with slightly higher humidity.   On the 27th, we pointed the telescope at the zenith and spun the azimuth axis over a number of points, covering 360$^\circ$ in 12$^\circ$ steps with three extra points overlapping.  Figure \ref{fig:sky89} shows the computed degree of polarization and figure \ref{fig:sky89ang} shows the position angle of polarization in the instrument frame.  The first measurement is at 292$^\circ$ at 4:16UT, wrapping around and finishing at 316$^\circ$ at 4:56UT.  Since the telescope never deviated from the zenith, the sinuosoidal oscillation in the degree of polarization is entirely caused by the telescope.  In the hour it took to make these measurements, the sun had dropped several degrees in altitude, increasing the overall degree of polarization at the zenith.  This is causing the points in the right side of the figure to not close the sine-curve.  The plot of position angle has been folded back on itself by 180$^\circ$ since the Stokes parameters are inherently 180$^\circ$ ambiguous. The position angle should go through two complete 90$^\circ$ rotations.  The position angle does this rotation, however it does not do it linearly!  The deviation is at least 30$^\circ$ from linear and the deviation is nearly identical over the hour of the measurement, in contrast to the 10\% change seen in the degree of polarization over the same time period.  Since the rotation of the plane of polarization matches so well over the entire hour of observations, we are confident that this rotation is entirely the telescope.  Also, during the hour this experiment was performed, the overall intensity of the sunlight decreased linearly by 50\% with no dependence on azimuth, showing that there is minimal absorption of different polarization states.  

 %%%%%%%%% Figure 32 here

\begin{figure}[!h,!t,!b]
\includegraphics[ width=0.8\linewidth, angle=90]{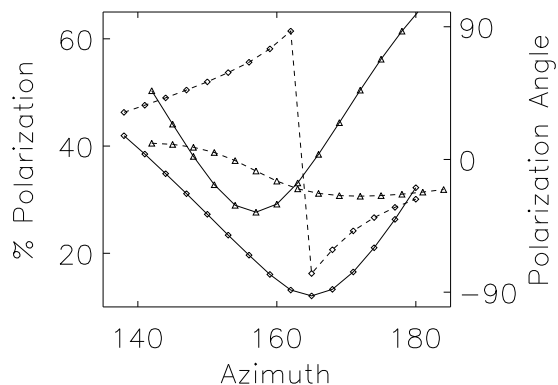}
\caption[sky89-depol]{\label{fig:sky89-depol}
The measured degree of polarization and position-angle of polarization of scattered sunlight at the zenith during sunset near the maximum depolarization azimuths.  The solid lines show the degree of polarization with the scale on the left.  The dashed lines show the instrumental position angle of polarization with the scale on the right.  The polarization measurements at azimuths of 232-274$^\circ$, shown in triangles, have been shifted by 90$^\circ$ to fit on this plot. }   
\end{figure}

	Another thing to notice is that the minimum polarization at 160$^\circ$ and 340$^\circ$ is close to 10\%, whereas the minimum at 70$^\circ$ and 250$^\circ$ is more like 30\%.  On the 28th, we did a more in-depth study of the polarization minima in order to more fully quantify the minima.  We observed azimuths of 138-180$^\circ$ and 232-274$^\circ$ in 3$^\circ$ steps spending 13 minutes at each group at 4:26UT and 4:40UT respectively.  Figure \ref{fig:sky89-depol} shows the degree of polarization and measured instrumental position angle for these observations.  The azimuth's have been shifted by 90$^\circ$ to center them.  The deep minimum shows a very strong PA flip through the minimum.  The shallow minimum shows a much weaker angle dependence.  

 %%%%%%%%% Figure 33 here

	In summary, the twilight observations show that the telescope can cause a 75\% linearly polarized source to depolarize to 15\%.  The telescope can also rotate the plane of polarization by at least 30$^\circ$ with pointing.  The telescope also causes the rotation of the plane of polarization with wavelength by up to 90$^\circ$ from 6500$\AA$ to 9500$\AA$.  All of these effects are strong functions of pointing.  The 2005 measurements suggest that the depolarizing effect greatly diminishes at longer wavelengths, but that the rotation of the plane of polarization continues.  These are very strong telescope polarization effects that must be considered when discussing results with this spectropolarimeter.

\subsection{Zemax Model of Telescope Polarization - Mueller Matricies}

	We constructed a model of the telescope's polarization response using Zemax to compare the predicted polarization effects with our observations.  Zemax software allows one to propagate any number of arbitrarily polarized rays of light through any optical design.  Each material or surface can be given any desired material property (thickness and complex index of refraction).  We wrote programs to compute the Mueller matrix of the telescope for any altitude or azimuth by tracking polarized light through our telescope with a range of mirror angles.  We traced the polarization properties of the telescope from the pupil to the first coud\'e focus, just after the coud\'{e} pickoff mirror (mirror 7).  This allows us to get a qualitative idea of the altitude-azimuth dependence of the polarization effects induced by the relative rotation of the altitude and azimuth mirror pairs.

 %%%%%%%%% Figure 34 here

\begin{figure}[!h,!t,!b]
\includegraphics[ width=0.8\linewidth, angle=90]{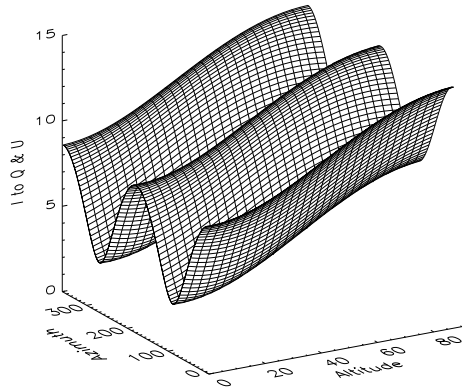}
\caption[zee-IP]{\label{fig:zee-IP}
The Zeemax calculated induced polarization, calculated as 100*(IQ$^2$+IU$^2$)$^{\frac{1}{2}}$ at the coud\'{e} focus for an index of refraction n=1.5-8i.  This shows the expected continuum polarization is of order 2\% to 15\% as a function of pointing.  A higher complex index reduces the degree of polarization, but does not significantly change the pointing dependence - all indicies give a double-sine with azimuth, rising with altitude.}   
\end{figure}

	We started by using the optical constants for aluminum from the Handbook of Optical Constants of Solids II (Palik 1991) and ended up testing a range of refractive indicies from 1.5-2.5 for the real part and 6-10 for the imaginary part.  These values are near the range of values measured for aluminum in the green to near infrared regions.  For instance, in the Handbook of optics the index goes from n+ik of (0.77, 6.1) to (1.47, 7.8) to (2.80, 8.5) for wavelengths of 500nm to 650nnm to 800nm respectively (Bass 1995).  Also, Giro et al. (2003) find an index of (0.64, 5.0) in a V filter (530nm) using a model of the Nasmyth focus of the alt-azimuth TNG telescope to fit observations of unpolarized standard stars.  This is entirely within the reasonable range.  
	
 %%%%%%%%% Figure 35 here

\begin{figure}[!h,!t,!b]
\includegraphics[ width=0.8\linewidth, angle=90]{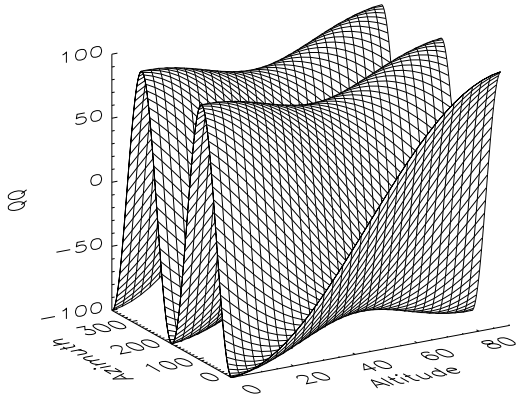}
\caption[zee-QQ]{\label{fig:zee-QQ}
The Zeemax QQ Mueller matrix term at the coud\'{e} focus for an index of refraction n=1.5-8i.  The matrix element has been multiplied by 100 for axis clarity.  This term shows how the overall reference frame of the instrument rotates with azimuth - a double-sine curve with azimuth is expected for a telescope with no polarization effects.  The slight bow seen at mid-altitudes reflects rotation of the plane of polarization by the telescope, causing small deviations from a perfect double-sine curve.}   
\end{figure}

	To compute the Mueller matrix, we input six pure polarization states $\pm$Q, $\pm$U, and $\pm$V at 400 pupil points each, propagate the light through the telescope at a given pointing, and average the resulting polarized light at the image plane.  Each reflection in the Zeemax models are given the same complex index of refraction.  The six results enable you to calculate the 16-element Mueller matrix for the telescope at a given pointing as the transfer from each input polarization to each output polarization.  These six calculations were computed for each pointing in 1$^\circ$ steps in altitude and azimuth mirror angles to project the Mueller matrix elements on to an all-sky map.  These maps were computed for a number of refractive indicies to determine the dependence of the telescope polarization on the assumed real and complex index components.  From these sky maps the expected amplitude and pointing-dependence of the telescopes polarization effects can be seen.  For notational simplicity, we will assume a Mueller matrix notation as in this equation:

\begin{equation}
{\bf M} =
  \left ( \begin{array}{rrrr}
  II   &   QI   &  UI   &  VI          \\
  IQ &  QQ  &  UQ &  VQ     \\
  IU &  QU  & UU  & VU         \\
  IV &  QV  &  UV  &  VV     \\ 
  \end{array} \right ) 
\end{equation}

	The complex component of the refractive index was the dominant term - the higher the complex index, the shallower the wave can penetrate the aluminum and hence the less significant the polarization effects.  We also ran a model with a complex index of 100, meaning essentially no aluminum penetration, and the models predicted perfect polarization response by the telescope.  The real component of the index had a much smaller effect, with higher indicies meaning more perfect polarization response, as expected.  
  	  
\begin{figure}[!h,!t,!b]
\includegraphics[ width=0.8\linewidth, angle=90]{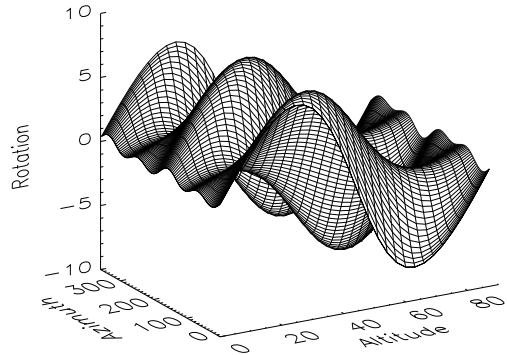}
\caption[zeerot]{\label{fig:zeerot}
The Zeemax calculated rotation angle in degrees at the coud\'{e} focus for an index of refraction n=1.5-8i.  The angle is calculated as the difference between the ideal telescope (with complex index -$\inf$) and the simulated telescope.  The angle is calculated as half the inverse tangent of QQ on QU.  }   
\end{figure}

	The telescope induced polarization from unpolarized incident light was computed as the IQ and IU Mueller matrix terms added in quadrature, 100*(IQ$^2$+IU$^2$)$^{\frac{1}{2}}$, shown in figure \ref{fig:zee-IP}.  The induced polarization is expected to be minimal when the mirrors are crossed and maximal when they are aligned.  The rotation of the plane of polarization can be calculated as the deviation of the frame rotation from the ideal frame rotation.  The ideal frame rotation is shown in figure \ref{fig:zee-QQ}.  This rotation angle is computed as $\frac{1}{2}atan(QQ/QU)$ with the large complex refractive index.  Figure \ref{fig:zeerot} shows an example of the telescopes rotation of the plane of polarization.  The rotation angles should be maximal at intermediate angles.  Since the azimuth axis has two angles where the mirrors are aligned (parallel and antiparallel), we expect a double-ridge structure in azimuth.   

 %%%%%%%%% Figure 36 here

  	These models proved ultimately to be quantitatively inaccurate.  Table \ref{zee} shows the range of Mueller matrix elements for a range of optical constants as well as the range of polarization-plane rotation angles and induced polarizations.  The predicted telescope-induced polarization (IQ and IU) were much greater, 2-28\% depending on pointing and index, than the measured 0.5\%-3.0\% polarization in unpolarized standards.  Even the double-sine shaped pointing dependence predicted by these induced-polarization models is masked by other effects.  
	
	The rotation of the plane of polarization was predicted to be quite small, 3$^\circ$ for a complex index of 10 to 8$^\circ$ for a complex index of 6 with an altitude-azimuth dependence illustrated in figure \ref{fig:zeerot}.  However, we observed very strong rotations, certainly at least 30$^\circ$, as previously seen in figures \ref{fig:skytheta} and \ref{fig:sky89ang}.  The model predicted the depolarization terms, QI UI and VI terms to be nearly identical to the induced polarization terms, IQ, IU and IV.  This predicted moderate to low depolarization, which clearly contradicts the twilight observations in figure \ref{fig:sky89}.  
	
	The rotation of the plane of polarization was also predicted to be only mildly dependent on wavelength since the complex index rises with wavelength from 6 at 500nm to near 11 at 1000nm to over 20 at 2000nm.  The deviation of the QQ, UU, QU and UQ terms from an ideal rotation matrix was small, at most 8$^\circ$ with a complex index of 6, and the rotation decreases with a rising complex index.  The indicies of refraction for aluminum do not change by much over this wavelength range.  In the Handbook of optics, the index of refraction goes from (1.4, 7.8) at 6500$\AA$ to (2.8, 8.5) at 8000$\AA$ and then to (1.2, 11.2) at 11300$\AA$ (Bass 1995).  The complex index of refraction is always 8 or higher, meaning the model will predict less than 4$^\circ$ rotation of the plane of polarization.  This also has been measured to be completely false by the twilight observations since the plane of polarization rotates by over 90$^\circ$ for some pointings, shown previously in figure \ref{fig:skytheta}.  

 %%%%%%%%% Table 2 here

\begin{table}[!h,!t,!b]
\begin{center}
\begin{tiny}
\hspace{5mm}
\caption{Zeemax Mueller Matrix Terms \label{zee}}
\hspace{-3mm}
\begin{tabular}{lcccc}
\hline
{\bf Mueller}              & {\bf Index}                   & {\bf Index}                    & {\bf Index}                       & {\bf Index}            \\                
{\bf Term  }               &{ \bf (1.5, 6)}                 & {\bf (1.5, 8)}                 & {\bf (1.5, 10)}                  & {\bf (2.5, 8)}              \\
\hline
IQ, QI                        & -0.17 to 0.28               & -0.10 to 0.17              &  -0.06 to 0.11                & -0.16 to 0.26           \\
IU, UI                        & -0.22 to 0.22               & -0.13 to 0.13              &  -0.09 to 0.09                & -0.20 to 0.20           \\
IV, VI                        & -0.04 to 0.08               & -0.02 to 0.03              &  -0.01 to 0.02                 & -0.03 to 0.05           \\
VQ, QV                    & -0.22 to 0.77               & -0.17 to 0.63              &  -0.14 to 0.53                 & -0.16 to 0.60           \\
VU, UV                    & -0.87 to 0.62               & -0.75 to 0.50              &  -0.64 to 0.41                 & -0.71 to 0.47           \\
VV                            & 0.41 to 1.00                & 0.64 to 1.00               &  0.76 to 1.00                  & 0.66 to 1.00             \\
Rot ($^\circ$)         & -8.1 to 7.6                   & -4.4 to 4.3                  &  -2.8 to 2.7                     & -4.2 to 3.7                 \\
\% Pol                     & 5.8 to 28.2                  & 3.4-16.4                     & 2.2 to 10.1                     & 5.2 to 25.5                 \\
\hline
\end{tabular}
\end{tiny}
\end{center}
\end{table}
		
\begin{figure}[!h,!t,!b]
\includegraphics[ width=0.8\linewidth, angle=90]{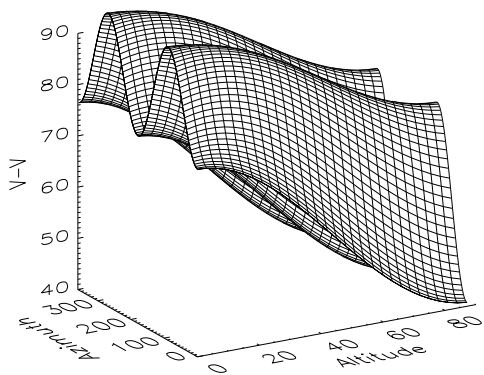}
\caption[zee-VV]{\label{fig:zee-VV}
The Zeemax calculated VV Mueller matrix term at the coud\'{e} focus for an index of refraction n=1.5-8i.  The matrix element has been multiplied by 100 for axis clarity.  This term shows a significant loss of circular polarization (V) to other polarization states at certain pointings. }   
\end{figure}

	We are unable to measure circular polarization with our current instrument configuration, but the circular-induced term (IV and VI) were predicted to be very small, less than 0.1 for the lowest complex index and less than 0.02 for the highest.  The circular cross-talk terms (VQ, VU, UV, QV) were all predicted to be very significant, running up to nearly 1 as a function of pointing and index.  This also implies that significant V sensitivity will be lost at certain pointings, as illustrated by the VV term being less than 50\% for some pointings in figure \ref{fig:zee-VV}.  
	
	There is also further evidence that the index of refraction is significantly different for this telescope at longer wavelengths.  Circular cross talk has been measured from other instruments using the AEOS telescope.  The Lyot project uses a near-infrared (JHK) coronographic imaging polarimeter mounted in a different coud\'{e} room (Oppenheimer et al. 2003).  This room changes the zero-point of the azimuth mirror, but does not change the optical path.  They also utilize the adaptive-optics system for the telescope, which adds several reflections just before the coud\'{e} pickoff and outputs a collimated beam instead of a converging beam.  The instrument first has to change the beam diameter by a number of oblique folds.  The polarization analysis is then performed in the collimated beam just after the coronograph with liquid-crystal waveplates and a Wollaston prism.  Despite the differences between our optics and wavelengths, they also found very significant Q, U, V crosstalk that was pointing dependent in H band (1600nm) images (Oppenheimer et al. 2008).  The total polarization $\sqrt{q^2+u^2+v^2}$ for AB Aurigae was constant over a range of pointings though the individual terms varied.  Since the complex index is predicted to be over 16 at this wavelength the Zeemax models would produce very small cross-talk even with several added mirrors.   

  	The aluminuum oxide coating that forms on every mirror is a major source of uncertainty.  Since we are unable to model the effect of aluminum oxide on the mirror surfaces, which always form, are significant, and possibly variable in time, the models are not expected to be quantitative.  The handbook of optics states that the oxide layers typically reduce the optical constants by 25\% in the IR, 10-15\% in the visible, and not much in the UV (Bass 1995).  This reduction is small compared to the range of input parameters used for this analysis, however, with so many mirrors, we suspect the contribution to be significant.  Furthermore, our model does not include any of the spectrographs common fore optics.  
	
	The polarization measurements of these common optics described above showed that the degree of polarization remains nearly identical to the input polarization and that there are no orientation-dependent effects.  Complete linear polarization at the coud\'{e} focus (via a linear polarizer at the calibration stage) is completely reproduced by the polarimeter.  This suggests that these common foreoptics are not that significant in their overall effect on the degree of polarization for any input angle.  However the measurement does not constrain an orientation-independent rotation (a PA offset).  The flat field measurements shows 3.6\% polarization, showing that the polarization induced by the common fore optics is also small, even though there are 8 reflections including the three highly-oblique image-rotators reflections between the calibration stage and the slit.  This also suggests that most of the severe polarization effects are caused by the telescope mirrors.  These Zeemax models are only an illustrative guide of what polarization the relative rotation of the mirrors can induce and what the pointing dependence may look like.     

 %%%%%%%%% Figure 37 here

	We had claimed in Harrington et al. 2006 that the Zeemax model gives justification for neglecting circular polarization.  The circular polarization terms IV and VI were predicted to be quite small, typically less than 0.05.  The VQ, QV, UV and VU terms, which describe incident circular polarization becoming measured linear polarization were much larger, typically 0-0.7.  Since most astronomical sources have nearly zero circular polarization and the V signatures across H$\alpha$ in our stars has been measured to be very small we concluded that neglecting circular polarization was justified (cf. Catala et al. 2007 or Wade et al. 2007).  We have now shown these models to be quantitatively inaccurate and are uncertain of the telescope-induced circular polarization.  However, even though these effects are expected to be severe, since we are observing line polarization, even severe circular crosstalk will not impart more than a continuum-offset to our polarization measurements since they have negligable wavelength dependence across a single spectral line.  The linear polarization detections we report are typically an order of magnitude greater than the circular signatures reported elsewhere (cf. Harrington \& Kuhn 2007 and Wade et al. 2007) and these cannot be the result of cross-talk.

\subsection{Implications for spectropolarimetry}

	The telescope polarization effects can be quite severe and complicated.  The effect on stellar spectropolarimetry is significant, but useful observations can still be performed with this instrument.  The telescope can not induce any polarization effects across a spectral line because the wavelength dependence of the induced polarization is quite weak at these wavelength scales.  This means that any observed spectropolarimetric signature at some wavelength in a spectral line must be real, and at least at the observed magnitude.  The measured signature could have been depolarized, rotated, translated, or be a cross-talk signature from the stars circular polarization, but the wavelength of the change in the line cannot be affected and the amplitude serves as a lower limit on the true spectropolarimetric effect.

 %%%%%%%%% Figure 38 here

\begin{figure}[!h,!t,!b]
\hspace{2mm}
\includegraphics[ width=0.75\linewidth, angle=90]{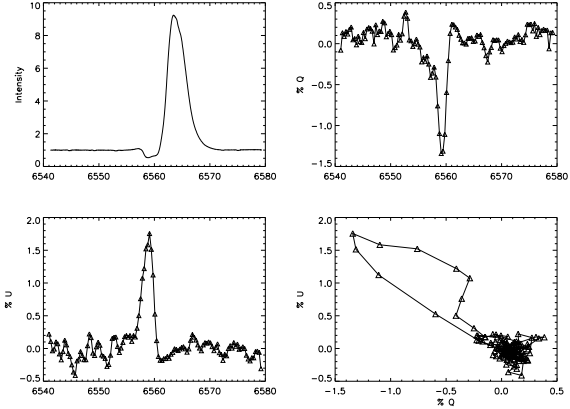}
\caption[iqu]{\label{fig:iqu}
A sample intensity, q, and u spectra of AB Aurigae, averaged heavily and plotted with wavelength and in the qu plane.  The top left box shows the intensity.  The top right and lower left show q and u respectively.  The bottom right shows a plot of q vs u, illustrating a qu-loop.  The continuum polarization has been set to zero, so a cluster of points representing this continuum for a knot around zero.  The loop is caused by the change of q and u across the absorptive component of the line.}
\end{figure}
	
 %%%%%%%%% Figure 39 here

\begin{figure}[!h,!t,!b]
\includegraphics[ width=0.8\linewidth, angle=90]{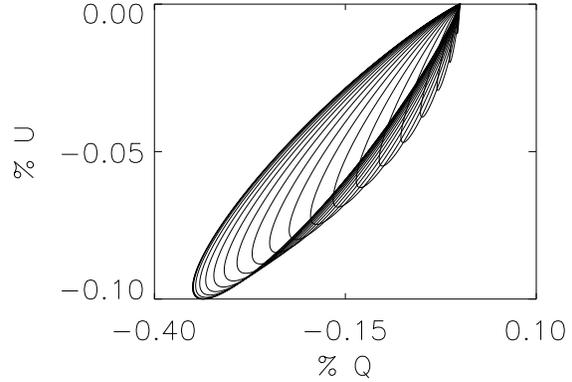}
\caption[telpol]{\label{fig:telpol}
A simulation showing how the telescope nullifies a qu-loop as it becomes an increasingly strong polarizer.  The telescope attenuates polarization at some direction until one polarization state is completely extinguished.  If the qu-loop is aligned with the position-angle of the polarizing telescope, the qu-loop is completel nullified.  However, to extinguish a qu-loop, the telescope must be nearly 100\% polarizing. }   
\end{figure}

	The effect of telescope polarization is best illustrated in the so-called qu-loops.  Polarization spectra are typically presented as either q and u spectra in iqu plots, or as degree of polarization and position angle spectra in ipt plots.  Some authors augment these with plots of the polarization in qu-space, ie plotting q vs u (cf. Vink et al. 2005b or Mottram et al, 2007).  When a polarization spectrum is plotted in qu space, where each wavelength becomes a point in q and u, the telescope's polarization effects can be simply represented as rotations, translations, shears, and amplitude-attenuations.  Figure \ref{fig:iqu} shows an example of a q and u spectrum of AB Aurigae plotted in the qu plane.  The spectra have been smoothed heavily to remove noise and clarify the example.  The overall intensity is plotted in the top left panel.  The q and u spectra, shown in the top right and bottom left respectively, show strong spectropolarimetric effects in the absorptive part of the line profile.  In the qu plane, when plotting q vs u, these spectra show a loop.  The polarization of the continuum outside the line creates the so-called continuum knot at zero.  Since the continuum polarization has been removed from these spectra, each point on the graph from outside the spectral line plots near zero forming a cluster of points.  The spectra show a change of -1.4\% and 1.7\% for q and u respectively, which occurs near the red side of the absorption trough.  These changes send the qu-loop off to at a PA of 135$^\circ$ to a maximum amplitude point (-1.4, 1.7).   
	
	Figure \ref{fig:telpol} shows an example of a telescope depolarization effect in qu-space.  A qu loop is shown with an initial amplitude of about 0.4\%.  With increasing telescope depolarizations, the telescope reduces the polarization of the source and the amplitude of a qu loop decreases.  Telescope-induced continuum polarization would cause these loops to cluster about some non-zero point.  Rotation of the plane of polarization could rotate this loop to an arbitrary angle.  In effect, the telescope can shift, reduce, or rotate a qu-loop, but it cannot create one.

\subsection{Calibration Summary}

	We have illustrated the polarization effects of a altitude azimuth telescope at the coud\`{e} focus.  The moving mirrors can induce, reduce, and rotate polarization as functions of position, pointing, and wavelength.  We measured these effects in a number of ways.  We used linear polarizers mounted at various points in the optical path to polarize the light and measure the output.  We used unpolarized standard stars as a method to measure the telescope-induced polarization.  We used twilight, a highly polarized source, to check the depolarization and rotation of the plane of polarization.  
	
	Our measurements with a linear polarizer mounted behind the spectropolarimeter showed that we can detect linear polarization downstream of the spectropolarimeter with an efficiency better than 95\% in the useful wavelength range of the linear polarizer.  The polarizer mounted just after the coud\'{e} pickoff mirror, in the calibration stage at the entrance of the room showed that the common fore optics produce a negligable depolarization for all linear polarization orientations.    	
	
	To do a complete calibration of the induced polarization, measurements of unpolarized sources have been made at many pointings.  The unpolarized star measurements have been projected back onto the sky so that a map of the telescope response has been made for the induced polarization to allow us to calibrate other sources of interest.  The unpolarized standard star measurements showed that the telescope induces polarization of order 0.5\%-3.0\% for the H$\alpha$ region in the new configuration.
	
	The scattered sunlight degree of polarization measurements showed very strong effects of the telescope on a highly polarized source.  The measurements at the zenith showed that the telescope can cause an effective depolarization, from 75\% to 15\% for the H$\alpha$ region.  This effective depolarization was seen to be much less severe at longer wavelengths in our 2005 observations, but we have not repeated those measurements with the new configuration.  The plane of polarization also rotates quite strongly.  The wavelength dependence of the rotation of the plane of polarization was also shown to be strong across all wavelengths, even as the depolarization effect decreased significantly. 
	  
	The telescopes polarization properties are quite severe and complicated.  The effect on spectropolarimetry is significant, but the telescope can not induce any effects across a spectral line.  In effect, the telescope can shift, reduce, or rotate a qu-loop, but it cannot create one.  Thus, any spectropolarimetric signature we detect is a useful lower limit on the magnitude of the polarization change across a line, and a good indicator of the wavelength range of the polarization effects.

\section{Instrument comparison: ESPaDOnS vs. HiVIS}

	As a verification and extension of our HiViS observations, we observed two targets with HiVIS and with another spectropolarimeter, ESPaDOnS, mounted on Mauna Kea's 3.6m Canada-France-Hawaii telescope (CFHT).  ESPaDOnS is a fiber-fed spectropolarimeter with a resolution of  68000, roughly 5 times higher than HiVIS in it's typical lowest-resolution polarimetric mode (Donati et al. 1999, Manset \& Donati 2003).  The spectropolarimeter has a completely different design with the polarizing optics mounted after the Cassegrain focus in a triplet-lens collimated beam.  The instrument uses a dual-beam design with three fresnel-rhomb retarders and a wollaston prism which produces the two orthogonally polarized beams.  The two polarized beams are then focused onto two fibers (with two additional for sky that are unused in polarimetric mode).  The typical exposure sequence is 8 exposures at different Frenel-rhomb orientations, 4 each for Q and U.  The telescope is equatorial and thus does not have any frame-rotation issues as the fibers are always at a fixed orientation on the sky.  This makes the instrument very useful for long integrations.  The fiber entrances do however introduce a strong instrumental continuum polarization, requiring a polynomial subtraction similar to HiVIS, and making continuum polarization studies impossible with this instrument.  This instrument has a very different setup from HiVIS and as such is a useful cross-check.  
	
 %%%%%%%%% Figure 40 here

\begin{figure}[htb]
\includegraphics[width=0.8\linewidth, angle=90]{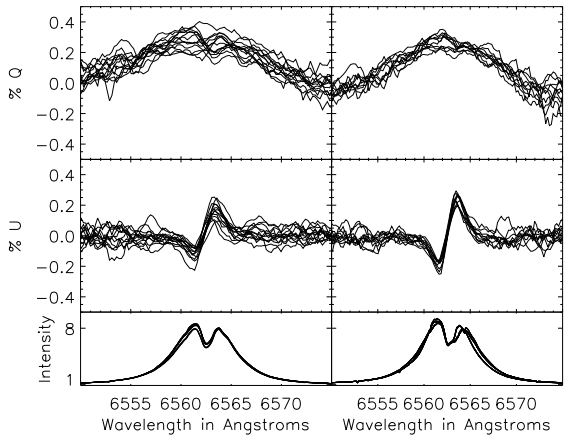}
\caption[cfh361]{\label{fig:cfh361}
The spectropolarimerty of MWC361 from HiVIS on the nights of June 18-21st 2007 is shown on the left and from ESPaDOnS on the nights of June 23 \& 24th 2007 on the right.  The MWC361 spectropolarimetry from HiVIS has been arbitrarily rotated by-eye to align the HiVIS qu-loops with those from ESPaDOnS.  Both data sets have their continuum polarization removed in the respective dedicated reduction scripts by low-order polynomial fits.  Both have been rebinned to lower spectral resolution for clarity.}
\end{figure}

 %%%%%%%%% Figure 41 here

\begin{figure}[htb]
\includegraphics[width=0.8\linewidth, angle=90]{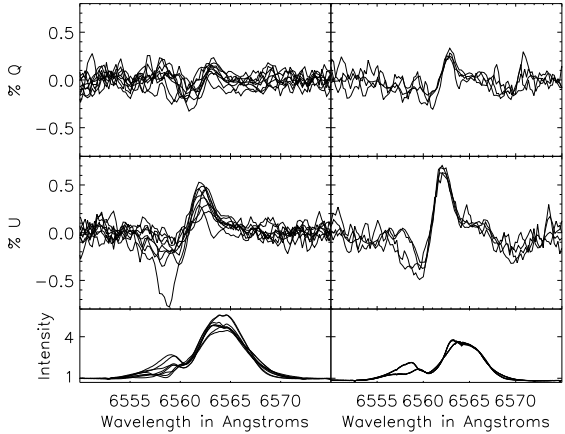}
\caption[cfh163]{\label{fig:cfh163}
The spectropolarimerty of HD163296 from HiVIS on the nights of June 18-21st 2007 is shown on the left and from ESPaDOnS on the nights of June 23 \&24th 2007 on the right.  No rotation has been applied to the HiVIS data.  The HiVIS data spans a range of pointings and exposure times.  There is a resemblance between some HiVIS and ESPaDOnS observations, but significant telescope polarization effects cause the amplitude of the HiVIS signature to be smaller.   Both data sets have their continuum polarization removed in the respective dedicated reduction scripts by low-order polynomial fits.  Both have been rebinned to lower spectral resolution for clarity.}
\end{figure}

	We observed MWC361 with the ESPaDOnS spectropolarimeter on CFHT on August 1$^{st}$ and 3$^{rd}$ of 2006 and with HiVIS on November 7$^{th}$ 2006.  We also have a back-to-back observation set of MWC361 and HD163296 with HiVIS on June 18-24th 2007 and with ESPaDOnS June 26th and 27th.  All data sets were taken in fair weather and reduced with the dedicated, automated Esprit data reduction package provided by CFHT (Donati et al. 1999).  As a further check, we adapted the HiVIS reduction package to process the ESPaDOnS data and found very similar results with both packages.  
	
	The ESPaDOnS data shows that over a period of three nights in 2006 (Aug 01 to Aug 03) and again over 2 nights in 2007 (June 23rd and 24th), the spectropolarimetry from MWC361 is identical within the noise.  The spectropolarimetry nearly a year apart matches very closely, though the intensity profiles change slightly.  Our HiVIS observations, while at lower spectral resolution, show a similar polarized spectrum.  A side-by-side comparison of the MWC361 data is shown in figure \ref{fig:cfh361}.  Each data set from HiVIS has been rotated by an arbitrary angle to take out any telescope frame rotation and plane of polarization rotations.  These by-eye rotations align the polarization spectra quite well, showing that HiVIS can reproduce the ESPaDOnS measurements quite well.   	

	The ESPaDOnS observations of HD163296 show a large amplitude polarization change, almost 1\% peak-to-peak, near the absorptive component of the emission line.  Observations with HiVIS also show a similar effect at these same wavelengths, but the magnitude of the change in the HiVIS data is smaller.  Figure \ref{fig:cfh163} shows the side-by-side comparsions of HiVIS and ESPaDOnS data.  The HiVIS data has also been de-rotated but the fits are not as good for many of the data sets.  This target had a different and wider range of pointings and could have the polarization amplitude modulated by the telescopes polarization properties.  Still, one can clearly see the proper qu shape and even similar values in the HiVIS observations.  

 %%%%%%%%% Figure 42 here

\begin{figure}[htb]
\includegraphics[width=0.7\linewidth, height=\linewidth, angle=90]{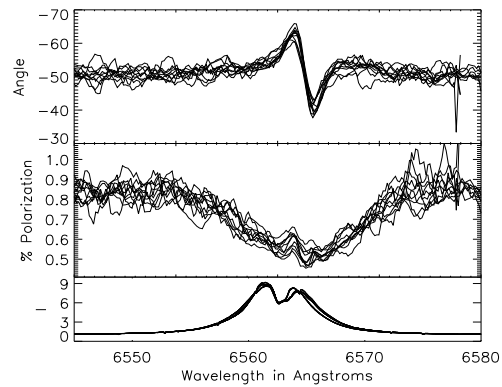}
\caption[cfht-vink]{\label{fig:cfht-vink}
The spectropolarimerty of MWC361 from CFHT on Aug 01 and 03 2006 plotted as degree-of-polarization and position angle.  The continuum polarization has been shifted to match the Mottram et al. 2007 polarization of 0.82\% at 95.6$^\circ$ or q-u continuum of (-0.80\%, -0.16\%) before calculating these polarization spectra.}
\end{figure}

	We can also compare our observations to those of MWC361 taken with the William Herschel Telescope (WHT) using the ISIS spectropolarimeter as presented in the Vink et al. 2002, 2005b and Mottram et al. 2007 papers.  The WHT is an alt-az telescope and ISIS is a long-slit spectropolarimeter.  The slit is aligned on the sky by physically rotating the entire spectrograph.  The polarizing optics are almost identical to the HiVIS design - a Savart plate just after the slit creates the to orthogonal polarizations with a rotating quarter- and half-wave retarder mounted in front of the slit.  A polarization observation is four sets of four-wave-plate rotations.  The slit however has only the primary and secondary mirrors upstream, and those near-normal reflections do not induce significant instrumental polarization (Harries et al. 1996).  However, the waveplate does produce a very significant 0.2\% amplitude ripple that must be subtracted from every high-resolution data set.  
	
 %%%%%%%%% Figure 43 here

\begin{figure}[htb]
\includegraphics[width=\linewidth, height=\linewidth, angle=90]{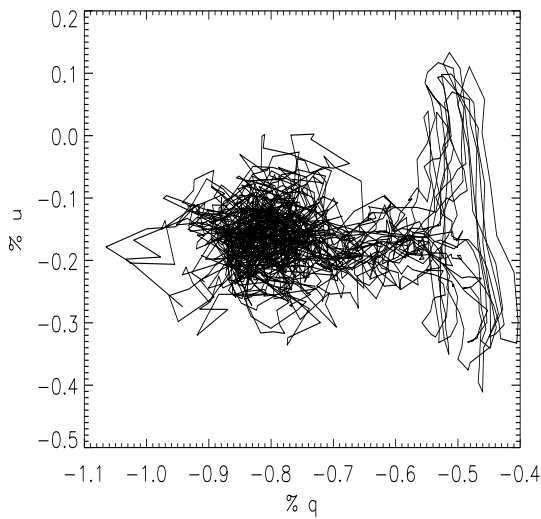}
\caption[cfht-loop]{\label{fig:cfht-loop}
The qu-loops from MWC361 from CFHT on Aug 01 and 03 2006 shown with the Mottram et al. continuum of -0.80\%, -0.16\%.  Stokes q decreases broadly over the line and u provides the small vertical deviations at the qu-loop's maximum amplitude.}
\end{figure}

	These WHT observations had a resolution of 35km/s or R$\sim$8500, whereas the 1.5" slit on the AEOS spectrograph has a resolution of R$\sim$13000 in the typical polarimetric mode.  All data sets are then binned by flux to a constant polarimetric error.  Vink et al. 2002, hereafter V02, present data from December 1999, Vink et al. 2005b, hereafter V05, presents data from December 2001 and Mottram et al. (2007) present data from late September 2004.  All three data sets show a similar qu-loop and polarization spectrum, though the Vink et al. 2002 data has much higher noise.  The continuum polarization was reported as 0.80\% to 0.82\% at a PA of 94$^\circ$ to 96$^\circ$ in V02 and V05 respectively.  We can take our CFHT and HiVIS data and scale them to these continuum values (q=-0.8\% u=-0.15\%) and calculate the degree of polarization and PA.  We find a very similar polarization spectrum, though with a significantly different PA spectrum, shown in figure \ref{fig:cfht-vink}.  This changing PA is caused almost entirely by the u spectrum changing sharply in the center of the line while q is positive over the whole broad region around the line.  In the V02 and V05 measurements, the PA change is on the red side of the emission line, with a roughly 20$^\circ$ change.  Our measurements now show a double-peaked change, still on the red side of the emission line with a similar wavelength range, but now with a two-fold change: a 15$^\circ$ increase in PA followed by a 15$^\circ$ decrease.  The significance of this is best seen in the qu-loops, shown in figure \ref{fig:cfht-loop}.  Our plot shows the u changes as both positive and negative with extreme values near -0.4\% and 0.1\% whereas the V05 qu-loops show only a significant decrease in u with a barely noticeable increase.  

	This illustrates the type of conclusions one can draw with HiVIS data.  MWC361 has a high declination, +68, and the range of pointings for all observations were quite small.  The star is always low and in the north, from altitudes of 30-42 and azimuths of 340 to 20.  HiVIS was quite successful at reproducing the spectropolarimetric measurements between ESPaDOnS and ISIS.  HD163296 however has a declination of -21 and spans a much greater range of pointings - altitudes from 30-48 and azimuths of 132-226.  The signature is still detected at the correct wavelengths, but at a reduced amplitude.
	
 %%%%%%%%% Figure 44 here

\begin{figure}[htb]
\includegraphics[width=\linewidth, height=\linewidth, angle=90]{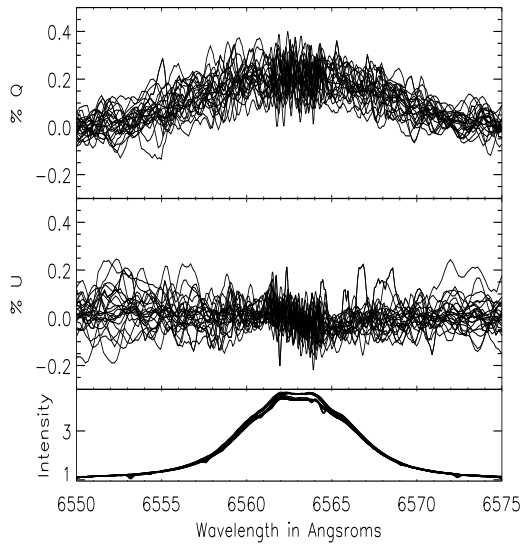}
\caption[gmcas]{\label{fig:gmcas}
Twenty-five q, u, and intensity spectra of $\gamma$ Cassiopeia taken on many nights from September 2006 to September 2007.  The plane of polarization has been rotated for each spectra to maximize q in the core of the emission line to provide a relative reference frame.  The average of all the polarization spectra is illustrated as the thick solid line showing a broad 0.2\% polarization change across the line.}
\end{figure}

	Another comparison between the William-Wehlau spectropolarimter or WWS and HiVIS is possible (Eversberg et al. 1998).  The WWS has a design very similar to ESPaDOnS, but using two achromatic quarter-wave plates instead of two half-wave and one quarter-wave Fresnel rhombs.  Eversberg et al. 1998 report commissioning observations of $\gamma$ Cas on the 1.6m Observatoire du mont M\'{e}gantic (OMM) telescope that reproduce the 0.2\% polarization change originally seen in a Pockels-cell polarimeter measurement with the University of Western Ontario 1.2m telescope (Poeckert \& Marlborough 1977).  We see a similar smooth and broad signature in $\gamma$ Cas as well, shown in figure \ref{fig:gmcas}.  The observations have all been rotated by an arbitrary angle to maximize the +q spectrum in the core of the emission line.  With the simple, broad form of the polarization signature, this has the effect of producing an effective de-rotation to a common polarimetric frame.  The resulting polarization spectra are averaged to produce our best estimate of the average polarization signature, shown as the thick line in figure \ref{fig:gmcas}.  This average signature shows an amplitude of nearly 0.2\% with a very smooth profile.
	
	In summary, HiVIS is capable of reproducing spectropolarimetric measurements made with many other instruments.  However, there are some complications that arise from the polarization properties of the coud\'{e} path that cause some spectropolarimetric measurements to be significantly reduced in magnitude and/or rotated.

\section{Conclusions}

	We have presented our dedicated IDL-based reduction package and the polarization properties of the HiVIS spectropolarimeter.  This package compares very favorably with other dedicated reduction packages.  The Esprit package on ESPaDOnS and our scripts give very similar results when applied to ESPaDOnS data.  We have found that the telescope induces 0.5\% to 3.0\% continuum polarization in an unpolarized source with a strong dependence on secondary focus.  The telescope can also strongly depolarize a source or rotate the plane of polarization as was shown by highly-polarized twilight observations.  All of these are functions of altitude, azimuth, and wavelength.  Optical ray trace models with Zeemax were presented to illustrate the pointing dependence of the telescope polarization properties, but were ultimately shown to be quantitatively inaccurate due to the very common problem of oxide coatings.  
	
	While these telescope effects complicate the analysis of data taken with this instrument, useful information about the amplitude and wavelength dependence of the polarization across a spectral line can still be obtained.  The performance of the instrument was compared to other spectropolarimeters, particularly ESPaDOnS on CFHT, ISIS on WHT, and WWS on OMM.  HiVIS found to reproduce the spectropolarimetric signatures, sometimes strikingly well given the possible complications.  This instrument will be able to make sensitive spectropolarimetric observations, despite some complications in the interpretation of the polarization after so many oblique reflections.  

\subsection{Acknowledgements}
This program was supported by the NSF AST-0123390 grant, the University of Hawaii and the AirForce Research Labs (AFRL).  This program made use of observations obtained at the Canada-France-Hawaii Telescope (CFHT) which is operated by the National Research Council of Canada, the Institut National des Sciences de l'Univers of the Centre National de la Recherche Scientifique of France, and the University of Hawaii.  This program also made use of the Simbad data base operated by CDS, Strasbourg, France.  The authors also wish to thank Don Mickey for many useful discussions about polarimetry and optical design.

\end{document}